\documentstyle[12pt]{article}

\begin{document}

\title{ \bf Combinatorial nuclear level density \\
by a Monte Carlo method }

\author{
Nicolas Cerf $^{a,b,}$\thanks{Research Assistant of the Belgian National Fund
for Scientific Research} \\
\\
{\normalsize $^a$  Institut d'Astronomie et d'Astrophysique, C.P. 165,} \\
{\normalsize 50, av. F.-D. Roosevelt,} \\
{\normalsize 1050 Bruxelles, Belgium.} \\
\\
{\normalsize $^b$ Division de Physique Th\'eorique\thanks{Unit\'e de Recherche
des Universit\'es Paris XI et Paris VI associ\'ee au C.N.R.S.}\ ,} \\
{\normalsize Institut de Physique Nucl\'eaire,} \\
{\normalsize Orsay Cedex 91406, France.} \\
}

\date{}
\maketitle

\vspace{4 cm}

\vspace{4 cm}
\noindent { Submitted to Phys. Rev. C  \hfill September 1993 }

\newpage

\begin{abstract}

We present a new combinatorial method for the calculation of the nuclear level
density. It is based on a Monte Carlo technique, in order to avoid a direct
counting procedure which is generally impracticable for high-A nuclei.
The Monte Carlo simulation, making use of the Metropolis
sampling scheme, allows a computationally fast estimate
of the level density for many fermion
systems in large shell model spaces.
We emphasize the advantages of this Monte Carlo
approach, particularly concerning the prediction of the spin and parity
distributions of the excited states, and compare our results with those
derived from a traditional combinatorial or a statistical method.
Such a Monte Carlo technique seems very promising to determine accurate
level densities in a large energy range for nuclear reaction calculations.

\end{abstract}

\newpage

\section{Introduction}

The density of nuclear levels provides information about the structure of
highly excited nuclei, and is also a basic quantity in nuclear reaction
theory. For many years, measurements of nuclear level densities have been
interpreted in the framework of an infinite Fermi gas model
(see e.g. \cite{bib_enge}), and most calculations
have employed the well-known Bethe formula \cite{bib_beth}
since it predicts a general trend in reasonable accordance with
experimental data (i.e., an approximately exponential increase of the level
density with both the excitation energy and the mass number).
The Bethe formula is based on the statistical model
(see e.g. \cite{bib_eric,bib_huiz}), and involves three
major assumptions: ({\it i}) the independent particles assumption,
allowing to write a nuclear partition function in a simple form using
single-particle energies, ({\it ii}) an equidistant spacing of the
single-particle states near the Fermi level, and ({\it iii}) the
saddle-point approximation used to calculate the inverse
Laplace transform of the partition function.
As a result of assumptions ({\it i}) and ({\it ii}),
this simple model includes neither odd-even effects
nor shell effects, and it has been pointed out in ref. \cite{bib_marg}
that assumption ({\it iii}) is not very good in the nuclear case.
Therefore, various phenomenological modifications of this simple formula
have been proposed for use in practical calculations to account for these
effects (see ref. \cite{bib_ilji} for a recent review).
These are mostly based on the ad-hoc hypothesis that the same
functional form of the energy dependence as in the equidistant-level case
is valid, some free parameter(s) being introduced in order to match
experimental results. In the Newton-Cameron shifted Fermi-gas model
\cite{bib_newton,bib_camero}, the odd-even effects are included by means of
a pairing energy shift. The effective excitation energy is reduced by the
conventional pairing energy for the odd-mass and even-even nuclei,
resulting in a lower level density.
The pairing energy is determined from experimental odd-even mass differences,
and the level density parameter $a$ is the only adjustable parameter.
This model is able to describe experiments in a narrow energy interval
around the neutron binding energy, most experimental data being extracted
from the neutron resonances, but it does not allow to extrapolate the value
of the level density in either the low- or the high-energy regions.
This means that shell effects are not properly taken into account.
\par

In order to solve this problem, Gilbert and Cameron \cite{bib_gilbert}
proposed another formula (the composed four-parameter formula) which combines
the shifted Fermi gas formula at high excitation energies with a constant
temperature formula (see e.g. \cite{bib_eric2}) for lower energies.
By fitting the four constants in both regions, experimental data may be
well reproduced. Another simple approach, the back-shifted Fermi gas
model was later proposed \cite{bib_gadioli,bib_vonach,bib_dilg,bib_voneg}
in order to account for shell effects. In this model, both the
energy shift and the level density parameter $a$ are considered as adjustable
parameters, which allows to obtain a reasonable fit to the experimental
level densities over a wider range of excitation energies.
Afterwards, some phenomenological methods have been proposed to correctly
describe the thermal damping of shell effects with increasing
excitation energy (e.g. \cite{bib_shellIgna,bib_shellKata,bib_shellVona}).
In those methods, the idea is to reproduce as well as possible
the energy dependence of the level
density parameter $a$ taken from microscopic calculations, in order
to give better absolute values of the nuclear level density.
\par

However, most of the above-mentioned
semiempirical approaches are based on various drastic
approximations, and their shortcomings at matching experiments
are often overcome only by parameter adjustments. Therefore, the quality
of the phenomenological prescriptions for a wide range of mass numbers
can not be taken as a support for their accuracy outside the narrow
regions of excitation energy ($\sim$~5-15~MeV) and angular momentum
($\sim$~0-5~$\hbar$) to which most of the experimental knowledge
of level densities is confined.
Numerical statistical calculations have also been developed to calculate
a more realistic level density using the single-particle level scheme of the
shell model \cite{bib_marg,bib_bloc}, and the BCS formalism has been
included \cite{bib_deco,bib_igna} in order to account more properly for
pairing effects. However, those microscopic calculations are still
depending on the assumptions of the statistical model, namely the
saddle-point approximation which is not very good at low energy.
Moreover, the spin and parity distributions raise unsolved questions
and call for improvement.
\par
The advent of high speed computers has made possible to use methods
of calculation which do not depend on closed formulae, such as the
combinatorial method \cite{bib_grov,bib_hill,bib_ford}.
The level density is then calculated numerically by performing an exhaustive
counting of the nuclear excited configurations.
It yields an exact - at least
in the independent-particle picture of the nucleus - level density, but is
very time consuming and becomes intractable at high excitation energies or
for large shell model spaces (i.e. for high-A nuclei). The pairing
interaction may be included in the calculations
by applying the BCS theory, but at the expense of a strongly
increased computation time.
More recently, combinatorial calculations have been carried out
to estimate level densities with fixed exciton numbers \cite{bib_herm},
for use in preequilibrium models.
\par

A second method of computing nuclear level densities directly from a set
of single-particle states has been proposed by Williams \cite{bib_will},
in order to avoid the direct enumeration and classification of states
which renders the combinatorial method generally cumbersome.
It is based on the repetitive use of recursion relations to expand
the grand partition function, but it does not rely on the saddle-point
approximation like the traditional statistical method.
This so-called recursive method
provides an exact calculation of level densities, but, compared with
the combinatorial approach, it has the advantage of requiring far
shorter computation time.
It has been been extended in ref. \cite{bib_jacq} in order to
describe the spin-parity distribution, and also to provide state densities
with fixed particle-hole number.
Unfortunately, it cannot treat residual interactions exactly,
at least in its classical form, so that
it is useful in the framework of a non-interacting Fermi gas model only.
Spectral distribution methods \cite{bib_french,bib_haq}
(based on moment techniques to calculate averaged distributions of the
Hamiltonian eigenvalues) have been used
to provide level densities taking residual
interactions into consideration. However, these methods are generally
limited to a low-order expansion, which is not very accurate for large
shell model spaces.
Thus, for different reasons, both the combinatorial
and the recursive methods are inappropriate to yield microscopic
level densities including residual interactions in a reasonable
computation time.
\par

In this paper, we propose a Monte Carlo method as an alternative to the
traditional combinatorial method.
We calculate the level densities of spherical nuclei by
means of a Monte Carlo simulation based on the Metropolis
sampling scheme \cite{bib_metr}. We show that this technique,
initially devised for the simulation of molecular systems,
can be successfully applied to the nuclear level density
problem. It yields an exact - save the statistical errors -
level density within a short cpu time, whatever the mass number
of the nucleus or the considered excitation energy.
We also use this Monte Carlo method to calculate
the spin and parity distributions of the excited levels.
In this first attempt, we restrict ourselves to the case of spherical
nuclei, but the method could be easily extended to deformed nuclei.
A detailed description of the Monte Carlo method is given in
Section~\ref{Method}.
In Section~\ref{Results}, we check the reliability of this Monte Carlo
procedure by comparing our results with those of a direct counting
and with microscopic statistical calculations.
Section~\ref{Pairing} is devoted to the inclusion of pairing interaction
in this model. Finally, we conclude in Section~\ref{Conclusion}.
The Monte Carlo implementation of the spin coupling and the statistical
errors originating from the method are discussed in the Appendices.
More detailed results and a comparison with experimental data are
presented in a second paper \cite{bib_cerfsuite}.
\par

\section{The Monte Carlo method}
\label{Method}

As noted first by van Lier and Uhlenbeck \cite{bib_vanlier},
the nuclear level density problem in its simplest form
(i.e. the uniform spacing model with one kind of particles)
is equivalent to a well-known combinatorial problem in number
theory, the partitioning of integers, first solved
by Euler in the eighteenth century.
Indeed, for a system of $N$ fermions, the number $M=gU$ represents the
number of quanta which are to be distributed among the $N$ fermions,
$g$ and $U$ being the single-particle state density and the excitation
energy, respectively.
Thus, the state density
\begin{equation}
\omega(U)=g\ p_N(M)
\end{equation}
is proportional to the number $p_N(M)$ of partitions of the integer $M$
into no more than $N$ positive parts (see~\cite{bib_bohning}).
Note that an approximate solution for Euler's problem was derived
by Hardy and Ramanujan in ref. \cite{bib_hardy}, giving
\begin{equation}
p_N(M) \simeq { \exp (\pi \sqrt{{2 \over 3}M})  \over  \sqrt{48} M } \ ,
\end{equation}
when $N \ge M$ (i.e., when the excitation energy is not sufficient to excite
a particle from the lowest orbit, which is clearly the case for an
infinite model).
This approximation is in fact equivalent to Bethe's formula
\begin{equation}
\omega(U) \simeq { \exp (2 \sqrt{aU})  \over  \sqrt{48} U } \ ,
\end{equation}
with the parameter $a={\pi^2 \over 6}g$.
As investigated by Euler, the exact $p_N(M)$ can be calculated
by means of the recursive relation \cite{bib_bohning}
\begin{equation}
p_N(M)=p_{N-1}(M)+p_N(M-N) \ .
\end{equation}
It is instructive to notice that this relation is of the same class
as Williams recursive relation \cite{bib_will}
for the calculation of level densities in the uniform spacing model.
However, the Williams method is also applicable
for a non-equidistant single-particle level scheme
by expressing the single-particle energies as multiple
of an energy unit (chosen approximately equal to the accuracy on the
energies, e.g. 0.01~MeV).
In this case, the level density problem comes to a more complex
problem known in computer science as the subset-sum problem,
for which the best algorithmic solution is a recursive one \cite{bib_garey}.
Unfortunately, as mentioned above,
the recursive method can not incorporate residual interactions,
so that recourse to a direct counting method cannot be avoided
when treating pairing interaction.
On the other hand, it is well known that a direct counting of all
the excited nuclear levels is a very tedious task, almost impracticable
on the currently available computers whenever one is interested in
medium mass nuclei at intermediate energies.
Moreover, a direct counting
procedure cannot easily treat the residual interaction in nuclei
because it is then exceedingly time-consuming.
Thus it is natural to resort to a Monte Carlo technique
in order to avoid this exhaustive counting of the excited levels.
Furthermore, it is well known that Monte Carlo methods
provide generally very efficient algorithms for solving
combinatorial problems (e.g. combinatorial optimisation problems solved by
simulated annealing \cite{bib_kirk,bib_cerny}).
This is the main reason why we investigate the use of a Monte Carlo technique
in the context of the nuclear level density problem,
and this idea has been first explored in ref. \cite{bib_cerf}.
This Monte Carlo method
becomes essential if the residual interactions are taken into account, since
the recursive method is intractable in this case.
For simplicity, however,
we consider the nucleus as a system of non-interacting fermions
in this Section. This choice will allow a comparison with a direct counting
in Section~\ref{Results}.
The inclusion of the residual interactions is the subject of
Section~\ref{Pairing}.
\par

\subsection {Principle of the Monte Carlo procedure}

We present in this Section a Monte Carlo method
based on the Metropolis sampling scheme which makes possible the
determination of a combinatorial level density without direct counting.
The Metropolis technique was devised for the simulation of molecular
systems \cite{bib_metr}; it has since been
widely applied in statistical mechanics
(e.g. to calculate thermal averages) \cite{bib_hamm,bib_bind,bib_alle}.
We show here that it is convenient for describing nuclear level densities
as well.
Note that Monte Carlo simulations related to the
level density problem have been performed in ref. \cite{bib_augu}
to obtain quantities of interest for the determination
of the parity distribution. However, that work was done in the framework
of the statistical model, which appears to be unable to treat correctly
the parity dependence of the level density \cite{bib_cerfpar}.
\par

The Monte Carlo simulation is based on a random sampling of a very
small fraction of the excited states in the considered range of
excitation energy \cite{bib_cerfthesis}.
The resulting sample is assumed to be representative of
the whole configuration space, in analogy with what is done when
estimating a multi-dimensional integral by a Monte Carlo procedure
(see e.g. \cite{bib_hamm}).
Since the state density of a system of non-interacting fermions
may be trivially expressed as
\begin{equation}
\omega(U,J,\pi)=\sum_C \delta(U-U_C)\ \delta_{J,J_C}\ \delta_{\pi,\pi_C} \ ,
\end{equation}
i.e. as a multi-dimensional summation in a discrete space of configurations
(each configuration $C$ having an excitation energy $U_C$, spin $J_C$,
and parity $\pi_C$), it can be efficiently obtained
by a Monte Carlo sampling procedure.
Note that each $N$-particle state is called here a configuration;
it corresponds to the set of occupation numbers for all the single-particle
states. (This is often
called a combination in combinatorial methods terminology).
It is clear that such a procedure cannot yield the whole ensemble of
$\delta$-spikes, but it is not important in practice, since only the
average state density in energy bins is needed.
Thus, the properties of the whole spectrum of excited states
(i.e., energy, spin, parity, or other
quantities) can be simply derived by extrapolating from the sample and
applying the appropriate scale factor.
This procedure provides an {\it exact} state density in the sense that
it is independent of the grand-canonical statistical formalism
(i.e., the energy and the number of neutrons or protons are not fixed
only on average, which is a well-known source of errors).
Of course, one has to keep in mind that the results are inherently affected
by a statistical error.
\par
A Monte Carlo estimator for the state density can be simply written as
\begin{equation}
\omega(U,J,\pi) \simeq S \  \omega_{\rm smp}(U,J,\pi) \ ,
\end{equation}
where $S$ is the scale factor (to be determined) and $\omega_{\rm smp}$
stands for the state density in the Monte Carlo sample.
One has
\begin{equation}
\omega_{\rm smp}(U,J,\pi)=\sum_{i=1}^N
         \delta(U-U_i)\ \delta_{J,J_i}\ \delta_{\pi,\pi_i} \ ,
\end{equation}
where the {$C_i$} are randomly (uniformly) sampled configurations
(of energy $U_i$, spin $J_i$ and parity $\pi_i$) in the whole space,
and $N$ is the number of configurations in the Monte Carlo sample.
The scale factor can be expressed as
\begin{equation}
S=(\sum_C 1)/N \ ,
\end{equation}
but it can not be calculated by this method, since the
summation is prohibitively long in our configuration space
($\sum_C 1$ is intractable).
It has to be determined by an independent method (see below).
\par

Such an approach
to the nuclear level density problem, introducing random excited
configurations, may be quite justified. Indeed, the probability theory
has already been applied with success in the statistical model
to determine the asymptotic gaussian-shaped spin distribution
and the parity distribution (see e.g. \cite{bib_eric}).
Furthermore, the
nuclear level density problem may be seen as the problem of exploring a
very large multidimensional space of configurations. A Monte Carlo simulation
is recognized to be the best technique for solving this problem, and is in
fact the only available one in many cases \cite{bib_hamm}.
Note that, although the method is exact in principle,
the resulting $\omega$ is inherently affected by a statistical error
which scales like $1/ \sqrt {N}$ (see Appendix~\ref{app_errors}).
The counterpart of this problem
is that the accuracy of the results can be imposed by choosing an
appropriate size $N$ for the sample, and {\it do not depend}
on the actual number of states in the considered energy range.
(For instance, a
rough estimate may be obtained within a very short computation time by
sampling only a few configurations.)
On the contrary, when doing a direct counting of the states, the whole
configuration space has to be explored, yielding an exact result
within a computation time which can be prohibitively long. There is no other
alternative. This fact constitutes the major advantage of a Monte Carlo
procedure.
\par

\subsection {The Metropolis sampling scheme}   \label{sect_Metropolis}

It is clear that a simple random sampling of the states
would be completely unefficient,
since the state density $\omega(U,J,\pi)$ increases exponentially
with excitation energy $U$. It is necessary in our simulation to hinder
the random sampling of the highly excited configurations which are by far
the most numerous, in order to achieve the same level of accuracy over
the whole considered energy range. Therefore, we have to apply an
importance sampling method (see e.g. \cite{bib_hamm}).
Let us introduce a (discrete) probability
distribution $P(C)$. The Monte Carlo evaluation of $\omega(U,J,\pi)$
becomes
\begin{equation}
\omega(U,J,\pi) \simeq {1 \over N} \cdot \sum_{i=1}^N
    { \delta(U-U_i)\ \delta_{J,J_i}\ \delta_{\pi,\pi_i}
    \over P(C_i) }  \ ,
\end{equation}
whenever each $C_i$ is randomly sampled with a probability $P(C_i)$.
The problem is that the probabilities $P(C_i)$ cannot be simulated
(neither calculated) because it is impossible to perform the normalization
in our configuration space. The Metropolis method
provides a solution for this problem by only referring to ratios
of probabilities, and makes thus an importance sampling procedure practicable.
\par
The Metropolis sampling scheme allows to produce any random variable in many
dimensions with a given probability distribution. It only requires the
ability to calculate a weight function (proportional to the probability)
for a given value of the integration variable. Let us define a weight
function $W(C)$ that represents in our problem
the {\it relative} (unnormalized) sampling probability of the configuration
$C$, i.e., $W(C) \propto P(C)$.
At this point, $W(C)$ is arbitrary, but we will show how to define it properly
in the following.
The Metropolis method is based on a guided random walk which proceeds through
configuration space according to a given matrix of transition probabilities
${\bf p}$. Each element $p(C_a \to C_b)$ of this matrix corresponds to the
probability of performing a step from the state (or configuration) $a$ to
the state $b$. The Metropolis method prescribes the choice of this stochastic
matrix ${\bf p}$ in order to obtain a random walk with a limit distribution
(i.e., the distribution of the frequencies with which each state occurs
when letting the random walk run to infinity) equal to a given $P(C)$:
\begin{equation} \label{equ_metropolis}
p(C_a \to C_b) = \left\{
\begin{array}{ll}
q(C_a \to C_b)
 & \quad {\rm if~~} W(C_b) \ge W(C_a)\ , \\
q(C_a \to C_b) \cdot {\displaystyle W(C_b) \over \displaystyle W(C_a)}
 & \quad {\rm if~~} W(C_b)<W(C_a)\ ,
\end{array} \right.
\end{equation}
where ${\bf q}$ is the so-called underlying matrix
of the random walk, and will be defined below.
In order to completely define the random walk process, the diagonal elements
of the stochastic matrix must be taken as
\begin{equation}
p(C_a \to C_a) = 1 - \sum_{b \neq a} p(C_a \to C_b) \ .
\end{equation}
This Metropolis prescription has been proved to be rigorous for
a discrete space \cite{bib_hamm},
which is just the case for our nuclear configuration space.
Indeed, it is easy to check, using equation~(\ref{equ_metropolis}), that
the detailed balance
$P(C_a)\ p(C_a \to C_b) = P(C_b)\ p(C_b \to C_a)$
is verified provided that ${\bf q}$ is a symmetric matrix.
This clearly implies that, conditionally on the ergodicity of the random walk,
the correct limit distribution is reached.
As already mentioned, expression~(\ref{equ_metropolis})
only refers to the weight function $W(C)$, and is thus tractable.
\par
{}From the definition of the stochastic matrix ${\bf p}$, it appears that
each step (from state $a$ to $b$) can be decomposed into two stages.
The first one, corresponding to the stochastic matrix ${\bf q}$,
consists in the random choice of a trial move. Thus, $q(C_a \to C_b)$
stands for the probability to perform a trial move from state $a$ to $b$.
Note that the Metropolis method requires ${\bf q}$ to be a symmetric
matrix, so that the reverse move from state $b$ to $a$ has the same
probability as the forward move. The next stage is the decision to
either accept or reject this trial move, the probability of accepting or
rejecting a trial move depending on the given weight function $W(C)$,
as shown in (\ref{equ_metropolis}). If the trial move is rejected,
the system stays in the same state, that is $b=a$.
\par
Now, forming a long enough random walk following this procedure,
we have a new Monte Carlo estimator for the state density
\begin{equation} \label{equ_statedens}
\omega(U,J,\pi) \simeq S \  \sum_{i=1}^N
    { \delta(U-U_i)\ \delta_{J,J_i}\ \delta_{\pi,\pi_i}
    \over W(C_i) }  \ ,
\end{equation}
where the $C_i$ are the random configurations along the random walk of length
$N$.
Here, the scale factor is defined by
\begin{equation}
S=\left( \sum_C W(C) \right) /N \ .
\end{equation}
The second factor in the right-hand side of equation (\ref{equ_statedens})
represents the (corrected) state density in the Monte Carlo sample,
i.e., $\omega_{\rm smp}(U,J,\pi)$.
Thus, each sampled configuration $C_i$ is given an importance $1/W(C_i)$
that is inversely proportional to its sampling probability $P(C_i)$.
\par

\subsection{Implementation of the Monte Carlo calculation}
     \label{sect_Implementation}

In our calculation, each state belonging to the random walk corresponds
to a configuration $C$, i.e., it is associated with a set of occupation
numbers $\{ n_k^{\nu}(C) ; n_k^{\pi}(C) \}$
for all the neutron and proton single-particle states.
Thus, the successive configurations are chosen randomly (according to the
stochastic matrix ${\bf p}$) as follows.
First, one selects a nucleon at random (either a neutron with probability
$N/A$ or a proton with probability $Z/A$) and moves it to a randomly chosen
available single-particle state. This generates a new configuration.
(This single-particle trial move corresponds to our definition for the
underlying matrix of the random walk ${\bf q}$,
and is the most common choice for the elementary move.
This sampling procedure is shown in ref. \cite{bib_fren} to be generally
more efficient than
collective moves, with several particles moved simultaneously.)
Then, one accepts or rejects this single-particle
trial move according to equation~(\ref{equ_metropolis}).
This acceptance criteria will thus depend on $W(C)$, which will be
in general a function of $U_C$, $J_C$, and $\pi_C$.
In our calculation, an energy dependence has proven to be sufficient, so that
we choose $W(C)$ as a function of $U_C$ only.
This weight function is arbitrary in principle: different choices give the
same result for an infinitely long random walk. However, a judicious choice
makes the method much more efficient than the simplest choice $W(C)=1$.
Indeed, we have to compensate for the rapid increase of the number of states
with energy by selecting a correspondingly decreasing weight function.
An appropriate choice would be to take $W(C)$ proportional to
the inverse of a Bethe-type law,
\begin{equation}
W(C)=(a~U_C)^{5/4}~\exp \left[ -2~(a~U_C)^{1/2} \right]  \ ,
\end{equation}
in which the parameter $a$ is adjusted to ensure a more or less uniform
distribution of the sampled states in the considered energy interval.
We took this definition of $W(C)$ in our previous work (see \cite{bib_cerf}),
for simplicity.
In this paper, we prefer another definition based on
a recursive model calculation. This will be discussed later on.
\par

Let us consider now the
energy, spin, and parity assignment for each randomly sampled configuration.
The excitation energy $U_C$ of each configuration $C$
(when assuming a system of non-interacting fermions) is simply given by
the sum of the energies of all the occupied single-particle states
\begin{equation}
U_C=\sum_k \epsilon_k^{\nu} \cdot n_k^{\nu}(C) +
\sum_k \epsilon_k^{\pi} \cdot n_k^{\pi}(C) \ -\ E_0 \ ,
\end{equation}
where the $n_k^{\nu,\pi}(C)$ are the occupation numbers
for the configuration $C$,
and the $\epsilon_k^{\nu,\pi}$ are the single-particle energies
for neutrons and protons.
The ground state energy is noted as $E_0$.
It is clear that, with our procedure, each excited level will be sampled
with a probability proportional to its statistical weight,
as expected. Suppose for instance
that we are interested in the level corresponding to a (1d$_{5/2}$)$^3$
configuration, having a degeneracy $d={6 \choose 3}=20$.
It will obviously be sampled with a probability
proportional to the number of possible arrangements
of 3 particles into 6 orbits, that is to $d$.
The possible inclusion of pairing energy using BCS theory
(removing this degeneracy) will be examined below.
\par

The spin $J_C$ of each configuration $C$ is chosen randomly
among all the possible spins of its associated excited level;
we choose the sampling probabilities of all the spins
according to their respective degeneracies. In order to
achieve this random choice, we proceed as follows. First, we determine
the total spin of identical nucleons in each subshell. In order to account
for the Pauli principle, we use the so-called Mayer-Jensen table
\cite{bib_hill,bib_maye} which yields the number of times each spin occurs.
Thus, the spin is chosen at random with a probability distribution
determined according to this table,
as explained in Appendix~\ref{app_spincoupl}.
Then, the spins of the different subshells
are coupled randomly, for either type of nucleons separately.
For this purpose, we use the relation
\begin{equation} \label{equ_spincoupl}
J=j_1+j_2-{\rm min}(d_1,d_2)
\end{equation}
for the coupling of two spins ($j_1$ and $j_2$) to a total spin $J$,
where $d_1$ and $d_2$ are (discrete) random variables uniformly distributed
in [0,$2j_1$] and [0,$2j_2$] respectively. It can be shown
(see Appendix~\ref{app_spincoupl}) that this relation
yields a total spin $J$ obeying the usual coupling rule of angular momentum
for independent particles, with adequate weights (i.e., the sampling
probability of each spin is proportional to its degeneracy).
Thus, relation~(\ref{equ_spincoupl}) is applied recursively
in order to couple the spins of all the subshells.
Finally, the total neutron and proton spins are coupled at random, also
using equation~(\ref{equ_spincoupl}), to give the resulting spin of
the configuration $J_C$.
Note that, when doing a direct counting of the levels, the whole ensemble
of possible spins (for each excited level) has to be enumerated. On the
contrary, the Monte Carlo determination of the spin distribution is
far simpler: one has just to choose one of these spins with the
appropriate probability, which is much less time-consuming.
\par

The parity of the sampled configurations
is also easily obtained by combining the
parities of all the occupied single-particle states. Thus, the resulting
parity $\pi_C$ of a configuration $C$ is given by
\begin{equation}
\pi_C = \prod_k \ [\pi_k^{\nu}] ^{ n_k^{\nu}(C)}
        \prod_k \ [\pi_k^{\pi}] ^{ n_k^{\pi}(C)} \ ,
\end{equation}
where the $\pi_k^{\nu,\pi}$ are the single-particle state parities.
\par

Finally, one has to determine the scale factor $S$ needed
to obtain an {\it absolute} state density. To this end, we calculate
the total number of states $N_{\rm rec}$ up to our maximum energy
$U_{\rm max}$ (the upper limit of the energy interval of interest)
by means of the Williams recursive method \cite{bib_will}:
\begin{equation}
N_{\rm rec} = \int_0^{U_{\rm max}} \omega_{\rm rec}(U)\ {\rm d}U \ .
\end{equation}
Thus, we have a constraint on the Monte Carlo cumulative
state density at energy $U_{\rm max}$,
\begin{equation}
N(U_{\rm max})
  = \int_0^{U_{\rm max}} \sum_J \sum_{\pi} \ \omega(U,J,\pi) \ {\rm d}U
  = N_{\rm rec} \ ,
\end{equation}
which gives obviously for the scale factor $S=N_{\rm rec}/N$.
This simple recursive method
can neither yield the spin-parity distribution, nor take into account
a residual interaction.
However, we just need it to
normalize our Monte Carlo sample state density $\omega_{\rm smp}(U,J,\pi)$.
Moreover, as it provides the whole state density curve as a function
of the energy, $\omega_{\rm rec}(U)$, we can use it in order to define
our weight function as explained in Appendix~\ref{app_errors}.
This choice for $W(C)$
ensures a more or less uniform distribution of the sampled states
in the considered energy interval. As a consequence, the resulting state
density $\omega(U,J,\pi)$ is more or less equally accurate in the whole
energy interval. Note that we reject the sampling of excited states above
the upper limit $U_{\rm max}$ by putting $W(C)=0$ for $U_C>U_{\rm max}$.
This recursive calculation of the state density takes
a very short computation time.
\par

For illustrative purposes, we have calculated the Monte Carlo
level density of $^{56}$Fe up to an excitation energy of 30 MeV,
using the realistic set
of single-particle levels derived from a spherical Hartree-Fock calculation
based on a Skyrme interaction \cite{bib_tond}.
In order to obtain a reasonable accuracy,
the size of the sample $N$ (which crucially determines
the computation time) is adopted equal to $10^6$.
Figure~\ref{fig_totdensityFe} shows the derived cumulative
state densities (i.e. the number of states up to a given energy $U$).
The solid line represents the total cumulative state density
(with both parities), while the positive- and negative-parity cumulative state
densities are represented by dashed and dash-dotted lines, respectively.
The curves show an overall behaviour
in agreement with the statistical results, i.e. an asymptotic parabolic
shape versus energy (in logarithmic scale). However,
superimposed small-scale oscillations are pronounced at low energies, and
disappear only at higher excitation.
We also notice that the classical assumption of equipartition
of both parities \cite{bib_eric} is only valid above $\sim 15$ MeV.
Note that we obtain a high degeneracy for the ground state
because of the neglect of residual interactions. Of course, this
degeneracy would be removed if residual interactions were included in our
model (see Section~\ref{Pairing}).
For information, we also plot in Figure~\ref{fig_totdensityFe}
the cumulative number of sampled states along the Monte Carlo simulation.
It is evident that the excited states have been sampled approximately
uniformly over the whole energy interval [0--30~MeV], as expected,
so that the resulting accuracy is roughly constant
(see Appendix~\ref{app_errors}).
\par

\section {Results}
\label {Results}

We will show in this Section that the Monte Carlo procedure
is very efficient for calculating an exact - at least in principle - level
density with spin and parity dependence, whatever the mass number
of the nucleus or the considered excitation energy.
Other quantities,
such as the fixed particle-hole number level densities, could also be
obtained by this method. In this respect,
the Monte Carlo method must be considered as a way to implement
the calculation of a combinatorial level density, replacing a direct counting
procedure when it is unfeasible (for large shell model spaces).
However, a direct counting of the levels can
be helpful when the number of excited levels is directly computable,
in which case a Monte Carlo sampling would introduce
undesirable statistical errors.
\par

In ref. \cite{bib_cerf}, we reported on the Monte Carlo derived level
density of $^{56}$Fe, $^{208}$Pb, and $^{140}$La.
We chose the same size for the sample, i.e., $N=10^6$,
which gave essentially the same accuracy in the three cases
although the fraction of the whole set of excited
states that is actually sampled is about $3\times 10^{-4}$ for $^{56}$Fe,
$10^{-7}$ for $^{208}$Pb, and only about $2\times 10^{-9}$ for $^{140}$La.
With such a choice, it is remarkable that the Monte Carlo simulation
takes about the same computation time for the three nuclei, although their
mass numbers are very different.
Some other cases can be found in ref. \cite{bib_cerfpelaghia}.
In the following, we will limit ourselves to check the method by comparing
with other traditional methods for calculating level densities.
New results will be discussed in greater details in ref. \cite{bib_cerfsuite}.
\par

\subsection {Comparison with a direct counting}
\label {sect_MCcounting}

In order to verify the reliability of our Monte Carlo procedure,
we have performed calculations of the level density and
spin-parity distribution via a direct counting of the levels
(see ref.~\cite{bib_pich}).
We have considered in particular the case of $^{56}$Fe up to 30~MeV,
neglecting the pairing interaction and using the single-particle
levels from ref.~\cite{bib_tond}.
Note that a combinatorial calculation at such excitation energies
is still feasible in this mass number region,
although at the cost of an impressive cpu time.
The total state density derived with the Monte Carlo method
could hardly be distinguished on Figure~\ref{fig_totdensityFe}
from the direct counting \cite{bib_pich},
so that we did not plot the latter.
In fact, the tiny difference is simply due to the
statistical noise (which can be made arbitrarily small by increasing
$N$).
Of course, the recursive method also gives identical results, but,
nevertheless, it is impracticable when considering residual interactions.
\par

Let us turn now to the spin dependence of the $^{56}$Fe level density.
In Figure~\ref{fig_spinFe}, we plot the spin distribution
of the levels in the energy interval [20--21~MeV]
derived by the Monte Carlo method,
along with the direct counting calculation \cite{bib_pich}.
The square symbols and the solid line represent, respectively,
the result of the Monte Carlo method and the direct counting.
The Monte Carlo spin distribution is obtained by putting
$W(C)=1$ from $U=0$ to the upper limit $U=21~$MeV. Thereby,
the states with high excitation energy are sampled preferentially,
and the accuracy on the spin distribution is increased.
The adopted sample size is $N=50\times 10^6$,
in order to get very good statistics (to have negligible statistical errors).
Of course, such an accuracy might be unnecessary for practical applications,
but our intention here is just to show that the Monte Carlo result
is asymptotically exact.
Indeed, both Monte Carlo and direct counting calculations
are in very good agreement, showing the validity
of the random spin coupling procedure explained in Section~\ref{Method}.
Note that, in this high energy domain, the number of excited levels
is as high as $2\times 10^6$, so that the spin distribution
is expected to be consistent with the asymptotic distribution
of the statistical model (see Figure~\ref{fig_gaussianFe}).
The positive- and negative-parity spin distributions obtained by
the Monte Carlo method are also shown in Figure~\ref{fig_spinFe}.
It appears that, in spite of this large number of levels,
the parity equipartition
is not achieved yet in the considered energy interval,
illustrating the need for an appropriate
treatment of parity (see \cite{bib_cerfpar}).
\par

In the conventional statistical model, the dependence of the state density
upon the angular momentum projection $M$
has the gaussian form \cite{bib_eric,bib_huiz}
\begin{equation}   \label{equ_MCgaussian}
\omega(U,M)=\omega(U) { \exp(-M^2/2\sigma^2) \over \sqrt{2\pi\sigma^2} } \ ,
\end{equation}
where $\sigma^2$ is the spin cut-off parameter, and
$\omega(U)$ stands for the total state density.
The resulting spin-dependent level density
can be written as \cite{bib_beth}
\begin{eqnarray}     \label{equ_MCspinasympt}
\rho(U,J) &=& \omega(U,M=J) - \omega(U,M=J+1)  \nonumber \\
 &\simeq& - \left[ {{\rm d} \over {\rm d}M} \
          \omega(U,M) \right]_{M=J+1/2}  \nonumber \\
 &\simeq& { 2J+1 \over 2 (2\pi)^{1/2} \sigma^3 }
          \exp \left[ -{ (J+1/2)^2 \over 2\sigma^2 } \right]
          \omega(U)   \ ,
\end{eqnarray}
where $J$ denotes the total angular momentum of the levels at excitation
energy $U$.
\par
We check in Figure~\ref{fig_gaussianFe}
that our Monte Carlo derived spin distribution for $^{56}$Fe in [8--10~MeV]
looks similar to the gaussian-shaped asymptotic distribution.
We plot the $M$-distribution
of the excited states (i.e., $\omega(U,M)$ as a function of $M$).
The agreement with the gaussian form (\ref{equ_MCgaussian})
is evident, as expected given the excitation energy.
We estimate an effective spin cut-off parameter $\sigma^2$
by calculating the variance of this $M$-distribution. This yields
the value $\sigma^2\simeq 18.8$, corresponding to a maximum in the
$J$-distribution at a spin around $\sigma\simeq 4$. The gaussian
distribution with this adopted value for $\sigma^2$ is plotted.
Figure~\ref{fig_gaussianFe} also shows the corresponding $J$-distribution
of the excited levels simulated by Monte Carlo,
as well as the asymptotic distribution (\ref{equ_MCspinasympt})
obtained with the same value of $\sigma^2$.
It proves that one can be confident
in the asymptotic result if one is concerned with an energy
interval containing a sufficiently large number of levels.
Indeed, the number of levels amounts to $8 \times 10^3$ in our case.
However, when considering an interval containing fewer levels,
the assumption of a gaussian distribution becomes very poor
(see e.g. \cite{bib_cerf,bib_cerfpelaghia}).
It is interesting to note that the discrepancy between the
curves is more pronounced when looking at the $J$-distribution,
showing that the latter is more sensitive to deviations from
asymptotic results.
\par

Finally, the evolution of the parity distribution of the excited levels
in $^{56}$Fe as a function of energy (up to 30~MeV)
is shown in Figure~\ref{fig_parFe}.
We plot, as a function of $U$,
the cumulative parity asymmetry defined by
\begin{equation}
A={N^+(U)-N^-(U) \over N^+(U)+N^-(U)} \ ,
\end{equation}
with $N^{\pm}(U) = \int^U \omega^{\pm}(U) \  {\rm d}U$.
The dotted line corresponds to the Monte Carlo results, whereas the
solid line is calculated by a direct counting of the levels \cite{bib_pich}.
Once more, both curves are almost indistinguishable.
This proves that the Monte Carlo method is a powerful alternative
to the traditional combinatorial method;
it reproduces the parity-asymmetry curve derived by direct counting
within an asymptotically small statistical error and
with a huge gain of cpu time.
Note that the parity equipartition, generally assumed to be achieved
at low energies, is only reached above $\sim 15$MeV.
This point is examined in greater detail in ref.~\cite{bib_cerfpar}.
\par

\subsection {Comparison with a microscopic statistical model}
\label {sect_MCstatistical}

In order to illustrate the reliability of our Monte Carlo procedure,
we have also compared our results with the predictions of a
numerical statistical model. This model, based on the traditional methods
of statistical mechanics, takes into account the
shell effects by using a realistic discrete single-particle spectrum
(see e.g. \cite{bib_huiz}).
The total energy $E$, the number of neutrons $N$ and protons $Z$,
and the entropy $S$ are determined
from the single-particle energies $\epsilon_k^{\nu}$ (for neutrons)
and $\epsilon_k^{\pi}$ (for protons) by the relations
\begin{eqnarray}
E &=& -{\partial \Omega \over \partial \beta}   \label{equ_MCthermo1} \ ,\\
N &=& {\partial \Omega \over \partial \alpha_N}   \ ,\\
Z &=& {\partial \Omega \over \partial \alpha_Z}   \ ,\\
S &=& \Omega-\alpha_N N -\alpha_Z Z +\beta E    \label{equ_MCthermo2} \ ,
\end{eqnarray}
with
\begin{equation}
\Omega \equiv \ln Z
      = \sum_k \ln [1+\exp (\alpha_N-\beta \epsilon_k^{\nu})]
      + \sum_k \ln [1+\exp (\alpha_Z-\beta \epsilon_k^{\pi})]  \ ,
\end{equation}
where $Z$ is the partition function, and $\alpha_N$, $\alpha_Z$,
$\beta=1/t$ are the Lagrange multipliers related, respectively, to the
number of neutrons, the number of protons, and the total energy.
One also defines the quantity $D$ as the $3\times 3$ determinant of
the second derivatives of $\Omega$ with respect to
$\alpha_N$, $\alpha_Z$, and $\beta$, evaluated at the saddle point.
All these quantities can be calculated numerically, and the resulting
state density for a two-component system is given by
\begin{equation}   \label{equ_MCstatistical2}
\omega(E,N,Z) = { \exp S \over (2\pi)^{3/2} D^{1/2} } \ .
\end{equation}
These calculations were performed using the statistical method
from ref.~\cite{bib_arno}, based on a realistic spectrum
of single-particle levels from ref.~\cite{bib_tond},
neglecting the pairing interaction.
\par

In Figure~\ref{fig_statFeLa}a, we show the state density for $^{56}$Fe
calculated by this statistical method for comparison with the
combinatorial (Monte Carlo) result. The solid line represents the
state density obtained with our Monte Carlo method, plotted in bins
of 100~keV, while the dashed line corresponds to the statistical
model calculation. Both curves were calculated using the same set of
single-particle levels. Note that the noisy behaviour of the Monte Carlo
state density at low energies comes from the fact that a lot
of bins are devoid of levels. Of course, the curve could be smoothed
by increasing the bin width, but the oscillations (due to the
discrete structure of the single-particle spectrum) would then be smeared out.
The agreement between statistical and combinatorial state densities
is reasonable, even if those oscillations disappear completely
in the statistical formalism. The statistical state density can thus be
viewed as an average of the combinatorial state density,
which is obviously a consequence of the grand-canonical approach
(the energy is only fixed on average).
Note that the statistical curve looks higher than the averaged combinatorial
curve, but this is simply due to the logarithmic scale.
It has to be stressed that the discrepancies between both methods
(ranging up to one order of magnitude) might be significant
e.g. when calculating reaction cross sections
with statistical or Monte Carlo level densities.
We also consider the case of a heavier odd-odd nucleus
in Figure~\ref{fig_statFeLa}b. We compare the combinatorial (Monte Carlo)
and statistical state densities for $^{140}$La (assumed to be approximately
spherical, and without pairing). The solid and the dashed line correspond,
respectively, the Monte Carlo and the statistical
state density. The latter curve also appears as an averaged combinatorial
state density, even if a little systematic error seems to be present.
A small statistical noise is still visible in the Monte Carlo state
density, related to the choice of the sample size $N=10^6$.
\par

Let us consider now the spin distribution
in this microscopic statistical model,
and in particular the evolution of the spin cut-off parameter with energy.
Within the framework of the statistical model, one can generalize the
formalism by adding a supplementary Lagrange multiplier $\gamma$
related to the angular momentum projection $M$ (see \cite{bib_huiz}).
It can be shown that, except for extremely large values of $M$, the
Lagrange multiplier $\gamma$ is sufficiently small to obtain a good
approximation by expanding the statistical expressions in powers of $\gamma$,
and by keeping the terms up to $\gamma^2$ only. The thermodynamic quantities
of interest can be written in terms of $\gamma$ as in
equations (\ref{equ_MCthermo1})-(\ref{equ_MCthermo2}).
By eliminating $\gamma$,
one gets an expression for the $M$-dependent state density
similar to equation~(\ref{equ_MCgaussian}),
\begin{equation}
\omega(E,N,Z,M)= \omega(E,N,Z)
           { \exp(-M^2/2\sigma^2) \over \sqrt{2\pi\sigma^2} }  \ ,
\end{equation}
where $\omega(E,N,Z)$ is the total state density defined in
equation~(\ref{equ_MCstatistical2}).
The spin cut-off parameter which determines the width of the $M$-distribution
is given by the expression:
\begin{equation}
\sigma^2 =
  {1\over 4} \sum_k (m_k^{\nu})^2 {\rm sech}^2 {1\over 2}
                  (\beta \epsilon_k^{\nu} -\alpha_N)
 +{1\over 4} \sum_k (m_k^{\pi})^2 {\rm sech}^2 {1\over 2}
                  (\beta \epsilon_k^{\pi} -\alpha_Z)   \ ,
\end{equation}
where $m_k^{\nu}$ and $m_k^{\pi}$ are the single-particle magnetic quantum
numbers for neutrons and protons, respectively.
The resulting spin-dependent level density $\rho(E,N,Z,J)$ can be
expressed as in equation~(\ref{equ_MCspinasympt}).
\par

In Figure~\ref{fig_statspinSnDyPb}, we compare the effective spin
cut-off parameters of three nuclei ($^{120}$Sn, $^{162}$Dy, and $^{208}$Pb)
derived from our combinatorial (Monte Carlo) method
and from a numerical statistical model
(ref.~\cite{bib_arno,bib_gori}). Both predictions make use of
the same Woods-Saxon single-particle level scheme.
The Monte Carlo spin cut-off is derived by calculating the variance of the
$M$-distribution, as previously.
As can be seen, the Monte Carlo method leads to a rather good
prediction of the energy-dependent effective spin cut-off parameter,
showing that one can be confident in the method.
Note that, even if this agreement for $\sigma^2$ is very satisfying,
the statistical assumption of a gaussian distribution is not always justified
(see \cite{bib_cerf,bib_cerfpelaghia}), so that our
Monte Carlo procedure can become essential in some cases.
\par

\section {Pairing interaction}
\label{Pairing}

In our previous work \cite{bib_cerf}, we restricted ourselves to the
Monte Carlo evaluation of the level density without the inclusion
of pairing force. In this Section,
we explore the possibility to account for the pairing interaction
in the Monte Carlo method.
The pairing interaction is usually taken into account in combinatorial
calculations by use of the BCS theory
(see ref.~\cite{bib_hill,bib_herm}). For example,
Hillman and Grover \cite{bib_hill} corrected their combinatorially
calculated level density for pairing forces by solving the BCS equations
(at zero temperature) for each excited configuration. They used two
so-called ``interacting odometers'' to scan the nuclear configurations
systematically, one for the unpaired nucleon configurations (determining
the spin distribution), the other one for the pair configurations.
They extended the ``blocking'' method developed in ref.~\cite{bib_wahl}
for treating odd-$A$ nuclei (with one unpaired nucleon) to the case of
configurations containing more than one unpaired nucleon: all orbital pairs
containing unpaired nucleons are blocked (i.e., made unavailable for pair
diffusion by being omitted from the BCS calculation). In their model,
the total energy of a (proton or neutron) configuration $C$ is given by
\begin{equation}
E_C=\sum_k {}' \ \epsilon_k + \sum_k {}'' \ 2 \epsilon_k v_{k,C}^2
    -\Delta_C^2 /G   \ ,
\end{equation}
where $G$ is the pairing strength parameter, and $v_{k,C}^2$
is calculated from
\begin{equation}
v_{k,C}^2 = {1\over 2} \left\{ 1- {  \epsilon_k - \lambda_C   \over
       [ (\epsilon_k - \lambda_C)^2 + \Delta_C^2 ]^{1/2}  } \right\} \ ,
\end{equation}
$\Delta_C$ and $\lambda_C$ being obtained by solving the pair of BCS equations
\begin{eqnarray}
&& \sum_k {}'' \ 2 v_{k,C}^2 = \eta_C  \ ,\\
&& \sum_k {}'' \ [ (\epsilon_k - \lambda_C)^2 + \Delta_C^2 ]^{-1/2} = 2/G \ .
\end{eqnarray}
Here, $\Delta_C$ and $\lambda_C$ are the gap parameter and Fermi energy,
respectively, while $\eta_C$ stands for the number of paired nucleons.
The summation $\sum '$ is made over orbitals containing unpaired nucleons,
while $\sum ''$ is made over orbital pairs in which there is not an
unpaired nucleon.
Note that, as the excitation energy increases,
more and more orbital pairs are blocked, so that only the trivial
solution $\Delta =0$ can exist in general above a certain energy.
This behaviour is in qualitative agreement with the
result of the temperature-dependent BCS theory \cite{bib_deco}.
\par
However, the model of Hillman and Grover has an inherent drawback
related to the configurations involving promoted pairs. Due to its
variational nature, the BCS method cannot treat a general configuration
of pairs, but only yields the energy of the particular
configuration where the pairs are all placed in the lowest available
(unblocked) orbitals. They accounted for this problem by adding to the BCS
energy the difference in single-particle energy of that particular
configuration of pairs and the one under consideration.
They claimed that, even if this is clearly a rough approximation,
the resulting error is not very serious because the fraction of levels
with pair excitation is small.
\par
The combinatorial model of Herman and Reffo \cite{bib_herm}
is slightly different,
although the pairing interaction is also included by applying the BCS
theory to each configuration. As they were interested in estimating
a level density with fixed exciton numbers, they simply neglected the
excitations involving promoted pairs. All the excited particles are taken
as noninteracting excitons, and the orbitals in which they are placed
are excluded from the BCS consideration, even if two particles occupy
time-reversed orbitals. They also justified this approximation by arguing
that the number of such configurations involving promoted pairs
is very small.
\par
In conclusion, it is our opinion that this problem
of promoted pairs is due to the {\it hybrid}
character of those methods. Indeed, BCS is inherently a statistical theory,
and is therefore incompatible with any combinatorial procedure for
estimating the level density. In fact, only a kind of ``combinatorial''
treatment of the pairing interaction would give a consistent solution
to the above problem. This is our purpose in ref.~\cite{bib_cerfmartin},
where we propose
a quantum Monte Carlo method to treat exactly the pairing force in nuclei.
This method in its present status is, however, limited to the properties
of the ground state, and is not directly applicable to the level density
problem. Thus, in this paper, we have to limit ourselves
to an approximate inclusion of pairing by use of the standard BCS theory,
close to what is done in ref.~\cite{bib_herm},
and we show it is feasible at essentially no cpu cost.
\par

For each sampled excited configuration,
we solve the BCS equations in order to obtain its pairing energy $P_C$.
We also use the so-called blocking method \cite{bib_hill,bib_wahl},
that is the orbits occupied by unpaired nucleons
are blocked (i.e., unavailable for pair diffusion).
However, we treat the problem of the promoted pairs
in a slightly different way:
first, the number of unpaired nucleons occupying the Fermi level is
taken equal to the seniority $s$ derived from our random spin coupling
algorithm, as explained in Appendix~\ref{app_spincoupl}.
This will ensure that the spin attribution of the low-lying levels is
coherent with the single-particle level scheme.
Second, all the excited particles (and created holes) on the other levels
are taken as noninteracting excitons, even if two particles (or holes) occupy
time-reversed orbitals,
in analogy with what is done in ref.~\cite{bib_herm}.
We think that this is the simplest approximate prescription
to properly construct the first excited states,
while neglecting excitations involving promoted pairs at higher energies.
Thus, the energy $U_C$ of each sampled configuration is corrected for pairing
and replaced by $U_C-P_C$ in equation~(\ref{equ_statedens}),
when computing the Monte Carlo state density:
\begin{equation}
\omega_P(U,J,\pi) \simeq S \  \sum_{i=1}^N
{ \delta(U-U_i+P_i)\ \delta_{J,J_i}\ \delta_{\pi,\pi_i} \over W(C_i)} \ ,
\end{equation}
where $P_i$ stands for the pairing energy of the configuration $C_i$
(that is the sum of the pairing energies for neutrons and protons).
Note that the energy used to calculate the weight function (see
Appendix~\ref{app_errors})
does not include this pairing term, so that the state density without pairing
can also be computed at the same time. This state density is useful since
it allows to perform
the normalization according to $\omega_{\rm rec}$, which do not incorporate
pairing energy, as already mentioned.
In fact, the Monte Carlo simulation proceeds as if there was no pairing
(the weight function is inversely proportional to $\omega_{\rm rec}$),
and then the actual sampling of the configurations corrects for
the effect of pairing.
\par

In order to illustrate the effect of pairing, we have considered
the isobars $^{142}$Nd (even-even) and $^{142}$Pr (odd-odd).
For both nuclei, we use the spherical single-particle level scheme
resulting from a Hartree-Fock + BCS calculation from ref.~\cite{bib_tond}.
We solve the BCS equations for all the excited configurations
with the values for the pairing strength parameters
$G_n=2.25/N^{0.7}~{\rm MeV}$ (for neutrons)
and $G_p=2.00/Z^{0.7}~{\rm MeV}$ (for protons)
derived in ref.~\cite{bib_tond} for the ground state.
In Figure~\ref{fig_pairingGdEu}, we plot the cumulative total state density
including pairing effects derived with our Monte Carlo method
for both $^{142}$Nd (solid line) and $^{142}$Pr (dash-dotted line),
as a function of the excitation energy $U$.
The effect of deformation is neglected here. The dashed
and the dotted lines represent the cumulative state density for both nuclei
in the absence of pairing. The latter curves clearly
illustrate the existence of shell effects: as $^{142}$Nd
is situated at a neutron shell closure, its level density is lower than
that of the neighbour nucleus $^{142}$Pr. In fact, given the similarity
between the corresponding single-particle level schemes,
this can be simply described as a Rosenzweig effect \cite{bib_rosen},
resulting in a shell-correction shift in the energy scale.
In addition,
one can see in Figure~\ref{fig_pairingGdEu} that, for both nuclei,
the pairing force is responsible for a shift of the level density curve to
the right, according to the classical phenomenological models
(see e.g.~\cite{bib_hill}).
Roughly speaking, this shift corresponds to the condensation energy of the
ground state due to pairing. Therefore it is clear that,
due to the even-even character of $^{142}$Nd, its
level density has to be shifted more than that of $^{142}$Pr.
(The condensation energy is found to be about 4.6~MeV for $^{142}$Nd,
and 3.3~MeV for $^{142}$Pr.) The difference
between both shifts is approximately equal to $\Delta_n+\Delta_p$,
with $\Delta_n$ and $\Delta_p$ being respectively
the gap parameter for neutrons and protons in $^{142}$Nd.
The Monte Carlo method is, of course, more accurate than a simple
phenomenological model since it yields a shift varying with energy,
and thus provides a supposedly more realistic level density.
\par
Note that the extra amount of cpu time needed for the
inclusion of pairing in our Monte Carlo procedure was found
to be only about 30-40\%. This fact seems surprising at first
glance, since it is well known that a combinatorial computation becomes
in general extremely time-consuming when treating residual interactions.
It originates from the fact that only about 1\% of the Metropolis trial moves
are accepted on average along our random walk, so that the extra
cpu time needed to compute the pairing energy (by solving BCS equations)
only concerns about 1\% of the configurations.
Accordingly, the energy of the remaining (99\%) configurations does not need
to be calculated, allowing a significant gain of computing time.
The latter configurations are necessary, however, to associate the
appropriate statistical weights to the former 1\% configurations
that are effectively sampled.
\par
Therefore, we think that a Monte Carlo evaluation of the level density
with inclusion of pairing is very promising. If the pairing force could
be accounted for in a consistent way (i.e., without the use of BCS theory),
it would be the only available method for solving exactly this problem.
It would not depend on the grand-canonical statistical formalism
(finite-temperature BCS equations~\cite{bib_deco})
and its associated difficulties,
such as the spurious phenomenon of a sharp superfluid-normal
phase transition in nuclei~\cite{bib_rho}.
The exact solution of this problem via a Monte Carlo procedure
is the subject of a future work.
\par

\section {Conclusion}
\label{Conclusion}

The Monte Carlo method is shown to be a reliable technique for calculating
nuclear level densities with a sufficient accuracy within computation times
that are much shorter than those demanded by traditional combinatorial methods
(i.e., a direct counting).
Also, owing to its rapidity, this method makes possible the estimate of the
shell-model uncertainties (i.e., those associated with the single-particle
level scheme).
In the absence of residual interactions, however, the
recursive method is exact as well and requires approximately the same amount
of computation time.
In any case, it is necessary to calculate the Monte Carlo scale factor
by the recursive method. Thus the Monte Carlo method really becomes
a powerful tool when studying nuclear level densities with pairing
interactions, as explored in Section~\ref{Pairing}.
It would even be possible to include neutron-proton interactions
in this method, which cannot be considered by any other method.
\par

The Monte Carlo approach also turns out to be worthwhile to predict
the excited levels spin and parity distributions.
Considering the crude approximations made to obtain those
distributions in the framework of the statistical model,
the proposed method is an interesting alternative to the usual
statistical methods used in nuclear reaction calculations.
Moreover, the technique outlined in this paper could be extended
to the case of deformed nuclei.
More detailed calculations and a confrontation
with experimental data are presented
in a subsequent article \cite{bib_cerfsuite}.
\par


\bigskip
\noindent {\Large \bf Acknowledgments} \par
\bigskip

\noindent
We are grateful to B. Pichon for having introduced us to this subject
and for useful comments.
We would like to thank F. Tondeur and S. Goriely
for providing us with helpful results.
We also express our gratitude to M. Arnould, O. Martin, M. Rayet, J. Treiner,
and J. Vanhorenbeeck for many helpful discussions.
Finally, we acknowledge G. Paulus for a careful reading of the manuscript.
This work has been partially supported by the Science Program of the
Commission of the European Communities SC1-0065.
\par


\bigskip
\appendix
\noindent {\Large \bf Appendices} \par

\section {Statistical errors}
\label {app_errors}

Due to the principle of the Monte Carlo method (i.e., the sampling
of a finite number of configurations), it is clear that a statistical
noise is superimposed on the derived total state density and the
spin-parity distribution. We will only consider here the case of the
total state density, but the same reasoning also applies to the calculation
of the statistical error on other quantities.
\par

The continued evolution of the random walk yields successive configurations
$C$ that are distributed according to the weight function $W(C)$.
This distribution is then sampled by repeated application of the
stochastic process $p(C_i \to C_{i+1})$ defined in
equation (\ref{equ_metropolis}). The resulting Monte Carlo estimator
for the total state density is thus given by
\begin{eqnarray}    \label{equ_averomega}
\omega(U) &=& \left( \sum_C W(C) \right) \
 {\rm E} \left[ {\delta(U-U_C) \over W(C)} \right]_{P(C)} \ , \nonumber \\
 &\simeq& S \ \sum_{i=1}^N {\delta(U-U_{C_i}) \over W(C_i)} \ ,
\end{eqnarray}
where $S=\sum_C W(C) /N$ is the scale factor,
and ${\rm E}[~]$ means the expectation value
for a random sampling of the configurations $C$ according to the probability
distribution $P(C) \propto W(C)$.
\par

As explained in Section~\ref{sect_Metropolis}, the idea of the importance
sampling procedure is to achieve a more or less uniform sampling
in the considered energy range.
Let us consider an energy bin of width $\Delta U$, situated at an excitation
energy $U$, containing a non-zero number of states. (If the bin is devoid
of states, it must simply be excluded from the following considerations.)
Let us call $\{ {\bar C} \}$ the ensemble of configurations belonging
to this bin (i.e., for which $U \le U_{\bar C} \le U+\Delta U$),
and let us give them the same weight ${\bar W}(U)$ for simplicity.
Assuming that the state density is constant inside this bin
and equal to ${\bar \omega}(U)$, the weight of that bin is thus
approximately ${\bar W}(U) {\bar \omega}(U) \Delta U$.
The condition of a uniform sampling imposes thus that
\begin{equation}  \label{equ_uniform}
{\bar W}(U) {\bar \omega}(U) \Delta U /
\sum_C W(C) \simeq \Delta U /U_{\rm max} \ ,
\end{equation}
where $U_{\rm max}$ is the upper limit of the considered energy interval.
In practice, since ${\bar \omega}(U)$ is unknown, it is replaced
by the derived state density from the recursive method.
Equation~(\ref{equ_uniform}) corresponds to our definition
of the weight function $W(C)$.
\par

According to equ.~(\ref{equ_averomega}), the estimator
for the number of states in that bin is obviously
\begin{eqnarray}
{\cal N}(U) &=& S \ \sum_{i=1}^N
{\Theta(U_{C_i}-U) \Theta(U+\Delta U-U_{C_i}) \over W(C_i)} \ , \nonumber \\
&=& \left( S/{\bar W}(U) \right) \ \sum_{i=1}^N
\Theta(U_{C_i}-U) \Theta(U+\Delta U-U_{C_i}) \ ,
\end{eqnarray}
where $\Theta(x)$ stands for the Heaviside step function.
By integrating (\ref{equ_averomega}) between $U$ and $U+\Delta U$,
the expectation value of ${\cal N}(U)$ is trivially given by
\begin{equation}
{\rm E}[{\cal N}(U)] = \int_U^{U+\Delta U} \omega(U) {\rm d}U
= {\bar \omega}(U) \Delta U \ .
\end{equation}
In order to estimate the statistical noise
on ${\cal N}(U)$, let us define first
the probability of choosing a random configuration in this bin
as ${\cal P}={\bar W}(U) {\bar \omega}(U) \Delta U / \sum_C W(C)
    \simeq \Delta U/U_{\rm max}$.
Assuming that the configurations are independent (see below),
one has for its variance
\begin{equation}
{\rm Var}[{\cal N}(U)] = \left( S/{\bar W}(U) \right)^2 N
                         {\cal P} (1-{\cal P}) \ .
\end{equation}
Thus, the (relative) statistical error on ${\cal N}(U)$ is simply
given by
\begin{equation}
\epsilon = \sqrt{ 1-{\cal P} \over N {\cal P} }
         \simeq \sqrt{1 \over N {\cal P} } \ .
\end{equation}
If the sampled state distribution is more or less uniform, ${\cal P}$
will be approximately equal to $1/B$, $B$ being the number of bins
in the energy interval. Thus, the (relative) statistical error
will be on the order of $\sqrt{B/N}$,
i.e. the inverse of the square root of the number
of states which have been sampled in that bin (whatever the actual number
of states in that bin).
\par

Finally, another fact has to be taken into account when estimating
the statistical error on various quantities.
Indeed, the successive configurations along the random walk are clearly
not statistically independent since each one is generated by moving
one particle at most from the previous one; that is, $C_{i+1}$
is likely to be in the neighbourhood of $C_i$ even if the configurations
are distributed properly as the walk becomes long.
As a result, the above expression of the variance is clearly underestimated.
This can be quantified
by calculating the auto-correlation function of some estimator,
e.g. the energy $U_C$. In practice, the auto-correlation length
$L_{\rm corr}$ can be computed as the number of random steps for which
this function becomes reasonably small. Then, when estimating
the statistical errors on any quantity, the number of sampled states
$N$ has to be replaced by $N/L_{\rm corr}$, measuring the number of
independent sampled configurations during the Monte Carlo run.
\par

\section {Implementation of the random spin coupling}
\label {app_spincoupl}

The calculation of the spin distribution is simplified since we restrict
the method to spherical nuclei here.
The orbital angular momentum $j$ becomes thus a good
quantum number, allowing us to define subshells. The number of levels having
a given total angular momentum $J$ for a particular configuration of subshells
can be calculated by finding the number of ways the components $j$'s can
couple to $J$ according to usual coupling rules. We show in {\it i)} how we
achieve this coupling in a random manner.
However, when we treat two (or more) identical nucleons in a same subshell,
the restrictions imposed by the Pauli exclusion principle for fermions
complicate this coupling, as explained in {\it ii)}.
Note that we do not have to construct explicitly the eigenfunctions
of $J^2$ by use of the Clebsch-Gordan or fractional parentage coefficients,
as the method only serves to determine the correct number of states
with their good spin.
As a matter of fact, we make an arbitrary correspondence
between the $|m_1 m_2 \cdots \rangle$ states and the $|JM \rangle$ states.
We simply choose at random the $|m_1 m_2 \cdots \rangle$ states
so that the corresponding $|JM \rangle$ states
are sampled with the correct statistical weights.
Thus, we do not calculate an intermediate $M$-dependent state density
$\omega(U,M)$ in order to deduce $\rho(U,J)$ through the usual relation
$\rho(U,J)=\omega(U,M=J)-\omega(U,M=J+1)$, as in ref. \cite{bib_ford}.
Rather, we calculate
directly the spin distribution, which allows to properly incorporate
pairing effects.\par

\bigskip
\noindent {\it i) Spin coupling for different orbitals:}\par

Let us show how to couple randomly the spins of the different subshells
(for neutrons or protons). As mentioned in Section~\ref{sect_Implementation},
we use the following relation for the coupling of two spins ($j_1$ and $j_2$)
to a total spin $J$:
\begin{equation} \label{equ_J(j1j2)}
J=j_1 +j_2 - {\rm min}(d_1,d_2)  \ ,
\end{equation}
where $d_1$ and $d_2$ are discrete random variables uniformly distributed
in $[0,2j_1]$ and $[0,2j_2]$, respectively.
This relation yields a total spin $J$ obeying the usual coupling rule of
angular momentum for independent particles, with adequate weights
(i.e., the sampling probability of each spin is proportional to its
degeneracy).
\par
In order to prove this, let us calculate the probability distribution
of the random variable $J$ defined by equation~(\ref{equ_J(j1j2)}).
Since $d_1$ and $d_2$  are uniformly distributed, they have the following
cumulative distribution functions:
\begin{eqnarray}
    \label{equ_defF1}
F_1(d_1) &=& \cases{
   {\displaystyle d_1+1 \over \displaystyle 2j_1+1}
                        & ~~~if $0 \le d_1 \le 2j_1$ , \cr
   1                    & ~~~if $d_1 > 2j_1$ , \cr  }
\\  \label{equ_defF2}
F_2(d_2) &=& \cases{
   {\displaystyle d_2+1 \over \displaystyle 2j_2+1}
                        & ~~~if $0 \le d_2 \le 2j_2$ , \cr
   1                    & ~~~if $d_2 > 2j_2$ . \cr  }
\end{eqnarray}
Let us define the random variable $d={\rm min}(d_1,d_2)$, and let us
determine its cumulative distribution function $F(d)$ using
\begin{eqnarray}   \label{equ_defF}
F(d) &=& {\rm Prob} \left[ {\rm min}(d_1,d_2) \le d \right]
         \nonumber \\
     &=& {\rm Prob} \left[ (d_1 \le d) {\rm ~or~} (d_2 \le d) \right]
         \nonumber \\
     &=& 1 - {\rm Prob} [d_1 > d] \ \cdot \ {\rm Prob} [d_2 > d]
         \nonumber \\
     &=& F_1(d)+F_2(d)-F_1(d)\cdot F_2(d)   \ ,
\end{eqnarray}
where we have used the fact that $d_1$ and $d_2$ are independent random
variables.
Thus, if $d$ reaches the maximum value of
either $d_1$ or $d_2$
(i.e., $2j_1$ or $2j_2$ respectively), one has $F_1(d)=1$ or $F_2(d)=1$,
so that
\begin{equation}
F(d) \equiv 1 \qquad {\rm if~~}d \ge {\rm min}(2j_1,2j_2) \ .
\end{equation}
This implies that the probability of having $d > {\rm min}(2j_1,2j_2)$
is equal to zero, so that $J$ cannot be less than
$j_1+j_2-{\rm min}(2j_1,2j_2) = |j_1-j_2|$, as expected.
Using equations~(\ref{equ_defF1}), (\ref{equ_defF2}) and (\ref{equ_defF})
we can calculate the probability distribution $f(d)$ of
the random variable $d$ as
\begin{eqnarray}
f(d) &=& F(d) - F(d-1)
           \nonumber \\
     &=& { 2(j_1+j_2-d) + 1 \over (2j_1+1) (2j_2+1) }
           \qquad {\rm with~~} 0 \le d \le {\rm min}(2j_1,2j_2)     \ .
\end{eqnarray}
Thus, the probability distribution of $J$ is simply given by
\begin{eqnarray}
\Phi (J) &=& f(j_1+j_2-J)
              \nonumber  \\
         &=& { (2J+1) \over (2j_1+1) (2j_2+1) }
             \qquad {\rm with~~} |j_1-j_2| \le J \le j_1+j_2  \ ,
\end{eqnarray}
which exactly corresponds (within a multiplicative factor)
to the statistical weight of the spin $J$.
Note that the denominator comes from the normalization condition
\begin{equation}
\sum_{J=|j_1-j_2|}^{j_1+j_2} \Phi (J) = 1
\end{equation}
of the probability distribution $\Phi (J)$.
This procedure can be obviously generalized for the coupling of
$n$ independent spins by applying relation~(\ref{equ_J(j1j2)}) recursively.
\par

\bigskip
\noindent {\it ii) Identical nucleons in a single orbital:}\par

For identical nucleons in a subshell, only those states are permitted
for which the Pauli exclusion principle is obeyed.
Thus, the Monte Carlo implementation of the spin coupling is not
straightforward, as for different orbitals (case {\it i}).
A rapid method
for handling  this problem is to put the required spin distribution
in a table. This table, constructed according
to the Mayer-Jensen table \cite{bib_hill,bib_maye},
yields the number of times each spin $J$ occurs for various configurations
$(j)^k$, where $j$ is the spin of the subshell and $k$ is the number of
identical nucleons occupying the subshell.\par

As an example, we present in Table~\ref{tab_spin} a list of the possible
angular momenta $J$ for an orbital $j=9/2$. Note that the multiplicity
of the states due to the degeneracy for $M$ is taken into account.
The states are labelled by the seniority quantum number $s$ (corresponding
to the number of unpaired particles in the orbital), with
\begin{eqnarray}
s &=& 0,2,4,\cdots {\rm min}(k,2j+1-k) \qquad {\rm for~}k{\rm ~even}
   \ , \nonumber \\
s &=& 1,3,5,\cdots {\rm min}(k,2j+1-k) \qquad {\rm for~}k{\rm ~odd}
   \ .
\end{eqnarray}
For instance, the various spins of the two-particles states ($k=2$)
are obtained by summing the rows $s=0$ and $s=2$ in Table~\ref{tab_spin}.
The last column in Table~\ref{tab_spin} shows the total multiplicity of
the states of seniority $s$.
Note that when the subshell is more than half filled (i.e., when $k>j+1/2$),
one applies the same procedure with the number of holes
$2j+1-k$.
Our Monte Carlo procedure to give the spin of a configuration
$(j=9/2)^2$ consists in choosing at random one of the values
$J$=0, 2, 4, 6, 8 with the corresponding probabilities
1/45, 5/45, 9/45, 13/45, 17/45.
The total weight (needed to normalize the probabilities) is simply
$1+44=45={10 \choose 2}$.
At the same time, the seniority $s$ is chosen equal 0 or 2 according
to the probabilities 1/45, 44/45.
The value of the seniority $s$ yields the number of unpaired particles
in the considered subshell, needed when treating the pairing
interaction (see Section~\ref{Pairing}).
The same table for different values of $j$ has been computed, and
an analog procedure is used to yield a (random) spin for all
the orbitals.
\par





\newpage
\noindent {\Large \bf Tables} \par
\bigskip

\begin{table}[h]
\begin{center}
{\footnotesize
\begin{tabular}{   c     ccccccccccccc     r  }
\hline
\hline
$J$ &0&1&2&3&4&5&6&7&8&9&10&11&12& Total \\
\hline
$s=0$ &1& & & &  &  &  &  &  &  &  &  &  &1 \\
$s=2$ & & &5& & 9&  &13&  &17&  &  &  &  &44 \\
$s=4$ &1& &5&7&18&11&26&15&17&19&21&  &25&165 \\
\hline
$J$ & 1/2 & 3/2 & 5/2 & 7/2 & 9/2 & 11/2
    & 13/2 & 15/2 & 17/2 & 19/2 & 21/2 & 23/2 & 25/2 & Total \\
\hline
$s=1$ & & & & &10&  &  &  &  &  &  &  &  &10 \\
$s=3$ & &4&6&8&10&12&14&16&18&  &22&  &  &110 \\
$s=5$ &2& &6&8&10&12&14&16&18&20&  &  &26&132 \\
\hline
\hline
\end{tabular}
}
\vspace{1cm}
\caption{List of the statistical weights of the possible states
for different values of spin $J$ and seniority $s$ in a $j=9/2$ subshell.
The upper panel corresponds to states with an even number of particles $k$
$(s=0,2,\cdots k)$, while the lower panel corresponds to states with
odd $k$ $(s=1,3,\cdots k)$. }
\label{tab_spin}
\end{center}
\end{table}



\newpage
\noindent {\Large \bf Figure captions} \par
\bigskip

   \begin{figure}[h]
   \vspace{1cm}
   \caption{ Cumulative state density for $^{56}$Fe.
The solid line represents the total
(with both parities) density, while the positive- and negative-parity
densities are represented by dashed and dash-dotted lines, respectively.
The dotted line corresponds to the cumulative number of sampled states.
The adopted size of the sample is $N=10^6$. }
   \label{fig_totdensityFe}
   \end{figure}

     \begin {figure}[h]
     \vspace {1cm}
     \caption { Spin distribution (normalized to 1) of the excited levels
in the energy interval [20--21~MeV] for $^{56}$Fe. The squares
represent the results of our Monte Carlo procedure, while the
solid line represents those of
a direct counting calculation \protect\cite{bib_pich}.
Both calculations were made using the same single-particle level
scheme \protect\cite{bib_tond}.
The triangles and the crosses correspond, respectively,
to the Monte Carlo positive- and negative-parity spin distribution.
A sample size $N=50\times 10^6$ is adopted. }
     \label {fig_spinFe}
     \end {figure}

     \begin {figure}[h]
     \vspace {1cm}
     \caption { Distribution of the excited state
angular momentum projection $M$ and of the excited level spin $J$
for $^{56}$Fe in the energy interval [8--10~MeV].
The squares correspond to $\omega(U,M)$ resulting from our Monte Carlo
calculation, while the solid line represents the gaussian distribution
from eq.~(\protect\ref{equ_MCgaussian}).
The triangles stand for the Monte Carlo derived $\rho(U,J)$,
while the dashed line represents the asymptotic distribution
from the statistical model (eq.~\protect\ref{equ_MCspinasympt}).
The derived value for the effective spin cut-off parameter
is $\sigma^2\simeq 18.8$. }
     \label {fig_gaussianFe}
     \end {figure}

      \begin {figure}[h]
      \vspace {1cm}
      \caption { Cumulative parity asymmetry $A$
as a function of excitation energy $U$
for $^{56}$Fe. The dotted line represents the result of our Monte Carlo
procedure, while the solid line represents
those of a direct counting \protect\cite{bib_pich}.
Note that, for the sake of emphasizing the small discrepancies
between both curves, we have slightly shifted the dotted curve to the right.
Both calculations were made using the same single-particle level
scheme \protect\cite{bib_tond}. }
      \label {fig_parFe}
      \end {figure}

      \begin {figure}[h]
      \vspace {1cm}
      \caption { State density for $^{56}$Fe (a) and $^{140}$La (b).
The solid line represents
the combinatorial state density obtained with our Monte Carlo method,
plotted in bins of 100~keV, while the dashed line corresponds to
the statistical model calculation \protect\cite{bib_arno} using the same
spectrum of single-particle levels (from \protect\cite{bib_tond}).
The statistical curve appears as an averaged combinatorial state density. }
      \label {fig_statFeLa}
      \end {figure}

      \begin {figure}[h]
      \vspace {1cm}
      \caption { Comparison of the effective spin cut-off parameter
$\sigma^2$ as a function of energy $U$, for $^{120}$Sn, $^{162}$Dy,
and $^{208}$Pb. The symbols represent the values derived from our
Monte Carlo method (each symbol corresponds to a Monte Carlo simulation),
while the solid lines stand for the predictions of the statistical model
\protect\cite{bib_arno}. Both evaluations have been obtained making use of
the same Woods-Saxon single-particle level scheme. This figure is taken
from \protect\cite{bib_gori}. }
      \label {fig_statspinSnDyPb}
      \end {figure}

      \begin{figure}[h]
      \vspace{1cm}
      \caption { Cumulative state density including pairing effects
derived with our Monte Carlo method for two isobar ($A=142$) nuclei,
as a function of excitation energy $U$.
The solid and the dash-dotted line correspond, respectively,
to the case of $^{142}$Nd and $^{142}$Pr,
while the dashed and the dotted lines represent
the cumulative state density for both nuclei in the absence of pairing.
We took the single-particle level schemes and the pairing strength parameters
from ref.~\protect\cite{bib_tond} (i.e., $G_n \simeq 0.10~{\rm MeV}$
and $G_p \simeq 0.11~{\rm MeV}$ for both nuclei). }
      \label{fig_pairingGdEu}
      \end{figure}


\newpage
\begin{thebibliography}{99}

\bibitem{bib_enge} C. A. Engelbrecht and J. R. Engelbrecht,
                   Ann. Phys. 207 (1991) 1.
\bibitem{bib_beth} H. A. Bethe, Rev. Mod. Phys. 9 (1937) 69.
\bibitem{bib_eric} T. Ericson, Adv. Phys. 9 (1960) 425.
\bibitem{bib_huiz} J. R. Huizenga and L. G. Moretto,
                   Ann. Rev. Nucl. Sci. 22 (1972) 427.
\bibitem{bib_marg} H. Margenau, Phys. Rev. 59 (1941) 627.
\bibitem{bib_ilji} A. S. Iljinov, M. V. Mebel, N. Bianchi, E. De~Sanctis,
      C. Guaraldo, V. Lucherini, V. Muccifora, E. Polli, A. R. Reolon,
      and P. Rossi, Nucl. Phys. A 543 (1992) 517.

\bibitem{bib_newton} T. D. Newton, Can. J. Phys. 34 (1956) 804.
\bibitem{bib_camero} A. G. W. Cameron, Can. J. Phys. 36 (1958) 1040.
\bibitem{bib_gilbert} A. Gilbert and A. G. W. Cameron,
                      Can. J. Phys. 43 (1965) 1446.
\bibitem{bib_eric2} T. Ericson, Nucl. Phys. 11 (1959) 481.

\bibitem{bib_gadioli} E. Gadioli and L. Zetta, Phys. Rev. 167 (1968) 1016.
\bibitem{bib_vonach} H. Vonach and M. Hille, Nucl. Phys. A 127 (1969) 289.
\bibitem{bib_dilg} W. Dilg, W. Schantl, H. Vonach and M. Uhl,
                   Nucl. Phys. A 217 (1973) 269.
\bibitem{bib_voneg} T. von Egidy, H. M. Schmidt and A. N. Bekhami,
                    Nucl. Phys. A481 (1988) 189.
\bibitem{bib_shellIgna} A. V. Ignatyuk, G. N. Smirenkin, and A. S. Tishin,
                   Sov. J. Nucl. Phys. 21 (1975) 255.
\bibitem{bib_shellKata} K. Kataria, V. S. Ramamurthy, and S. S. Kapoor,
                   Phys. Rev. C 18 (1978) 549.
\bibitem{bib_shellVona} H. Vonach, M. Uhl, B. Strohmaier, B. W. Smith,
      E. G. Bilpuch, and G. E. Mitchell, Phys. Rev. C 38 (1988) 2541.

\bibitem{bib_bloc} C. Bloch, Phys. Rev. 93 (1954) 1094.
\bibitem{bib_deco} P. Decowski, W. Grochulski, A. Marcinkowski,
                   K. Siwek and Z. Wilhelmi, Nucl. Phys. A 110 (1968) 129.
\bibitem{bib_igna} A. V. Ignatyuk and Y. N. Shubin,
                   Sov. J. Nucl. Phys. 8 (1969) 660.

\bibitem{bib_grov} J. R. Grover, Phys. Rev. 157 (1967) 832.
\bibitem{bib_hill} M. Hillman and J. R. Grover, Phys. Rev. 185 (1969) 1303;\\
                   M. Hillman, Phys. Rev. C 9 (1974) 289.
\bibitem{bib_ford} G. P. Ford, Nucl. Sci. and Eng. 66 (1978) 334.
\bibitem{bib_herm} M. Herman and G. Reffo, Phys. Rev. C 36 (1987) 1546; \\
                   M. Herman, G. Reffo and R. A. Rego,
                   Phys. Rev. C 37 (1988) 797.

\bibitem{bib_will} F. C. Williams, Nucl. Phys. A 133 (1969) 33;\\
                   F. C. Williams, G. Chan and J. R. Huizenga,
                     Nucl. Phys. A 187 (1972) 225.
\bibitem{bib_jacq} C. Jacquemin and S. K. Kataria, Z. Phys. A 324 (1986) 261.

\bibitem{bib_french} J. B. French and K. F. Ratcliff,
                        Phys. Rev. C 3 (1971) 94;\\
                     F. S. Chang, J. B. French, and T. H. Thio,
                        Ann. Phys. 66 (1971) 137.
\bibitem{bib_haq} R. U. Haq and S. S. M. Wong, Phys. Let. 93B (1980) 357.

\bibitem{bib_metr} N. Metropolis, A. W. Rosenbluth, M. N. Rosenbluth,
                   A. H. Teller and E. Teller, J. Chem. Phys. 21 (1953) 1087.
\bibitem{bib_cerfsuite} N. Cerf, {\it in preparation}.


\bibitem{bib_vanlier} C. van Lier and G. E. Uhlenbeck, Physica 4 (1937) 531.
\bibitem{bib_bohning} M. B\" ohning, Nucl. Phys. A 152 (1970) 529.
\bibitem{bib_hardy} G. H. Hardy and S. Ramanujan, Proc. London Math. Soc.
                    Series 2, 17 (1918) 75.

\bibitem{bib_garey} M. R. Garey and D. S. Johnson, Computers
    and Intractability : A Guide to the Theory of NP-Completeness
    (Freeman, San Francisco, 1979).

\bibitem{bib_kirk} S. Kirkpatrick, C. Gelatt, and M. Vecchi,
                   Science 220 (1983) 671.
\bibitem{bib_cerny} V. Cerny, J. Optimis. Theor. and Appl. 45 (1985) 41.

\bibitem{bib_cerf} N. Cerf, Phys. Lett. B 268 (1991) 317.

\bibitem{bib_hamm} J. M. Hammersley and D. C. Handscomb,
                   Monte Carlo methods (Methuen, London, 1964).
\bibitem{bib_bind} K. Binder in: Monte Carlo Methods in Statistical Physics,
                   ed. K. Binder (Springer-Verlag, New York/Berlin, 1979).
\bibitem{bib_alle} M. P. Allen and D. J. Tildesley,
                   Computer Simulation of Liquids (Clarendon, Oxford, 1987).

\bibitem{bib_augu} W. Augustyniak, Z. Phys. A 317 (1984) 225.

\bibitem{bib_cerfpar} N. Cerf, Nucl. Phys. A 554 (1993) 85.

\bibitem{bib_cerfthesis} N. Cerf, Ph. D. Thesis (1993), {\it unpublished}.

\bibitem{bib_fren} D. Frenkel, in: Monte Carlo Simulations,
                   NATO-ASI Summer School, Bath (Catlow Editor, 1988).

\bibitem{bib_maye} M. G. Mayer and J. H. D. Jensen, Elementary
                   Theory of Nuclear Shell Structure (Wiley,
                   New York, 1955).

\bibitem{bib_tond} F. Tondeur, Nucl. Phys. A 303 (1978) 185;\\
                   F. Tondeur, Nucl. Phys. A 311 (1978) 51;\\
                   F. Tondeur, Nucl. Phys. A 315 (1979) 353.

\bibitem{bib_cerfpelaghia} N. Cerf, to appear in: Proc.
      Int. Conf. on the Future of Nuclear Spectroscopy,
      (Crete, 1993).

\bibitem{bib_pich} B. Pichon, Nucl. Phys. A (1993), {\it in press}.

\bibitem{bib_arno} M. Arnould and F. Tondeur, in: Proc. 4th
      Int. Conf. on Nuclei far from Stability,
      Helsing\o r, Eds. P.G. Hansen and O. B. Nielsen,
      (CERN 81-09, 1981), Vol.~1, p.~229.
\bibitem{bib_gori} S. Goriely, Ph. D. Thesis (1991), {\it unpublished}.

\bibitem{bib_wahl} S. Wahlborn, Nucl. Phys. 37 (1962) 554.
\bibitem{bib_cerfmartin} N. Cerf and O. Martin, Phys. Rev. C 47 (1993) 2610;\\
                         N. Cerf, Nucl. Phys. A (1993), {\it in press}.
\bibitem{bib_rosen} N. Rosenzweig, Phys. Rev. 108 (1957) 817;\\
            P. B. Kahn and N. Rosenzweig, Phys. Rev. 187 (1969) 1193.
\bibitem{bib_rho} M. Rho and J. O. Rasmussen, Phys. Rev. 135 (1964) B1295.

\end{thebibliography}
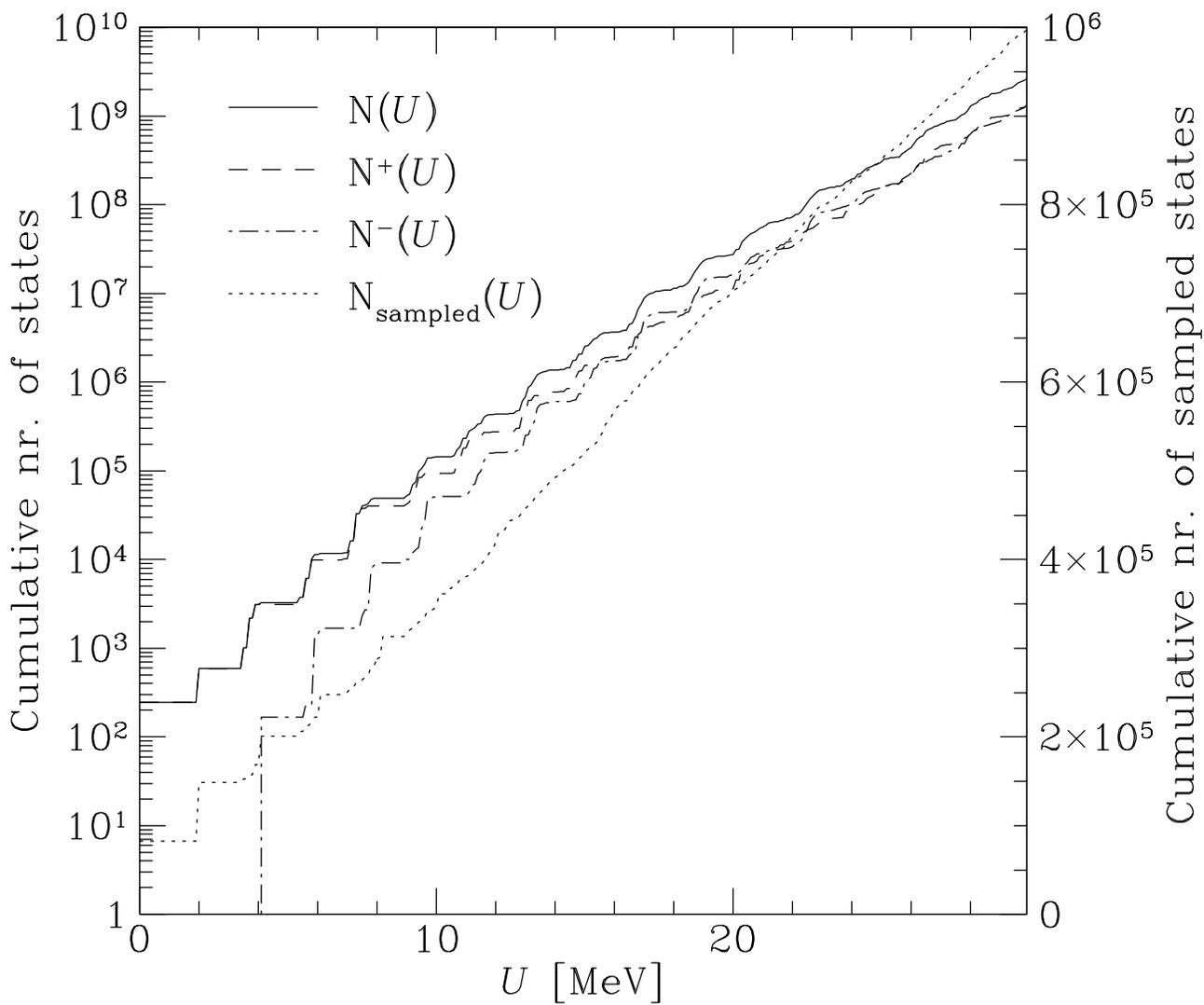
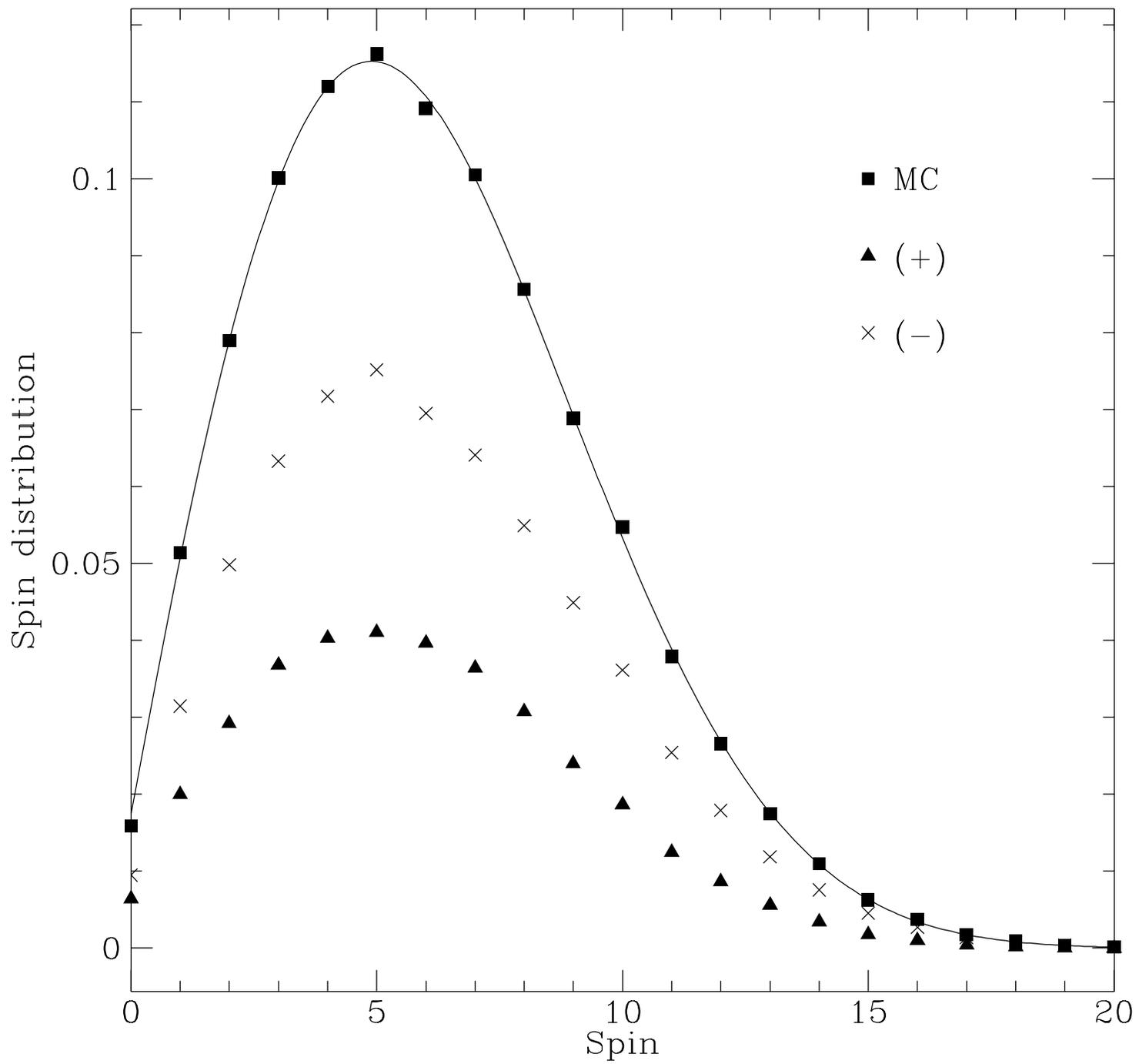
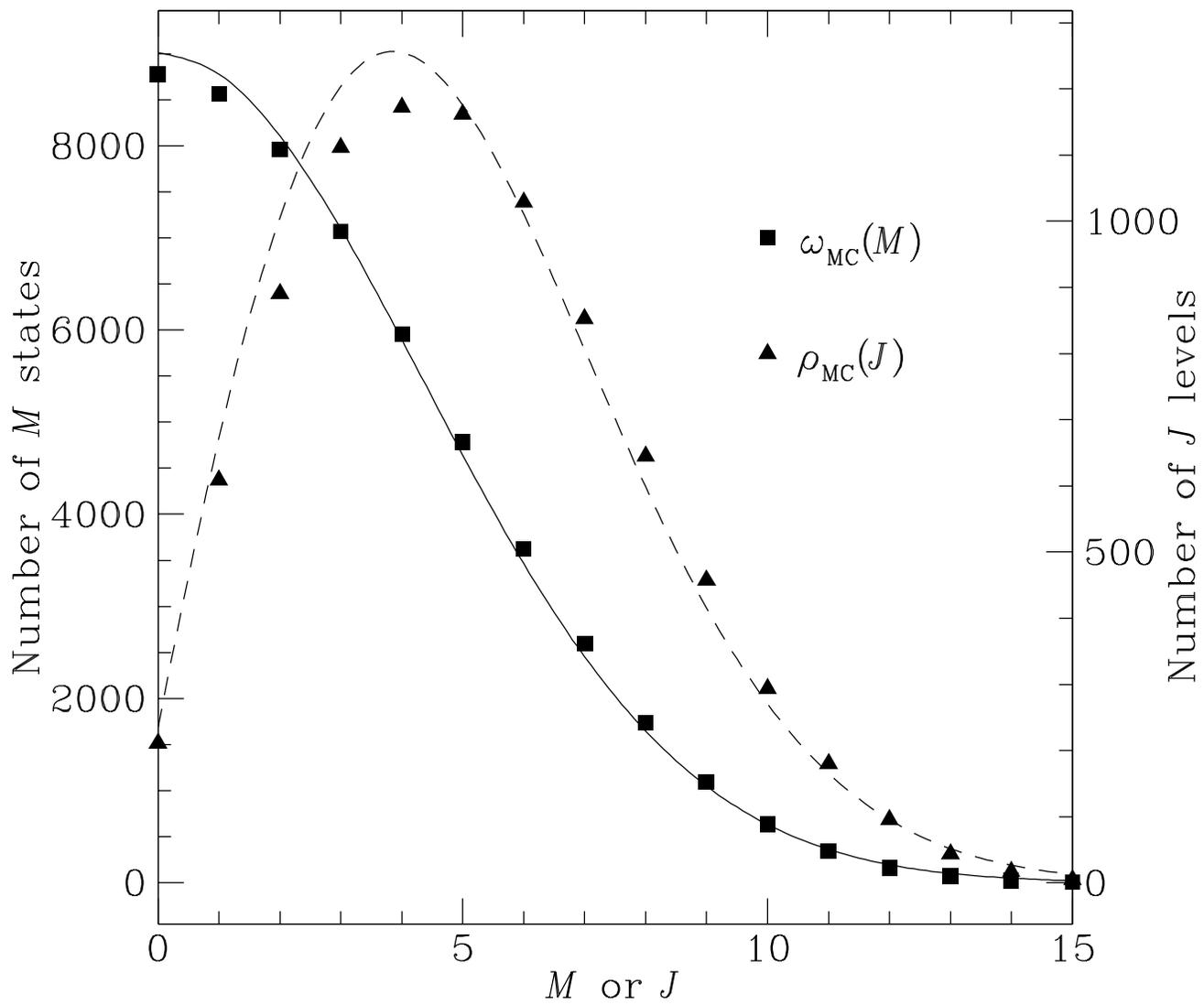
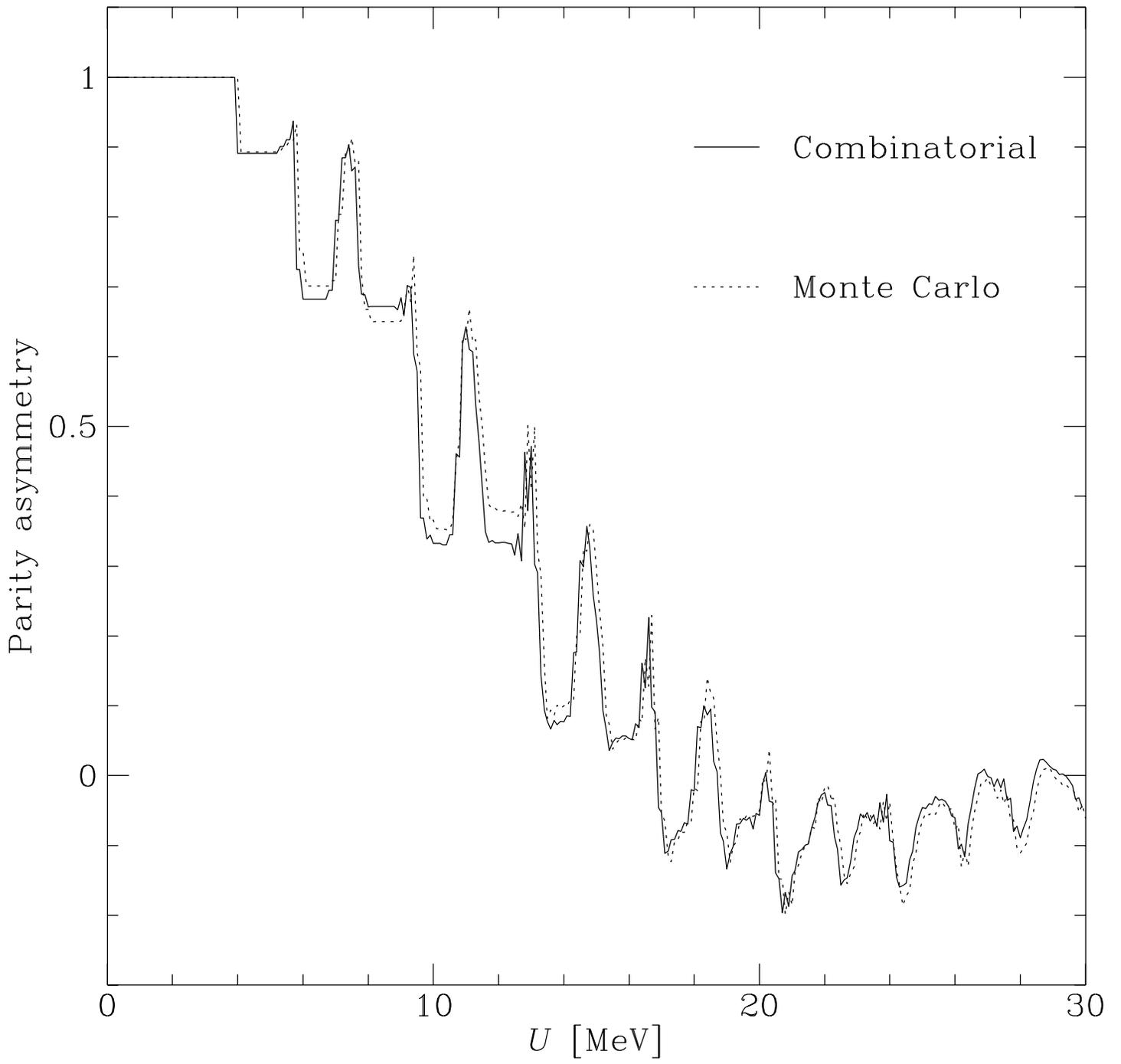
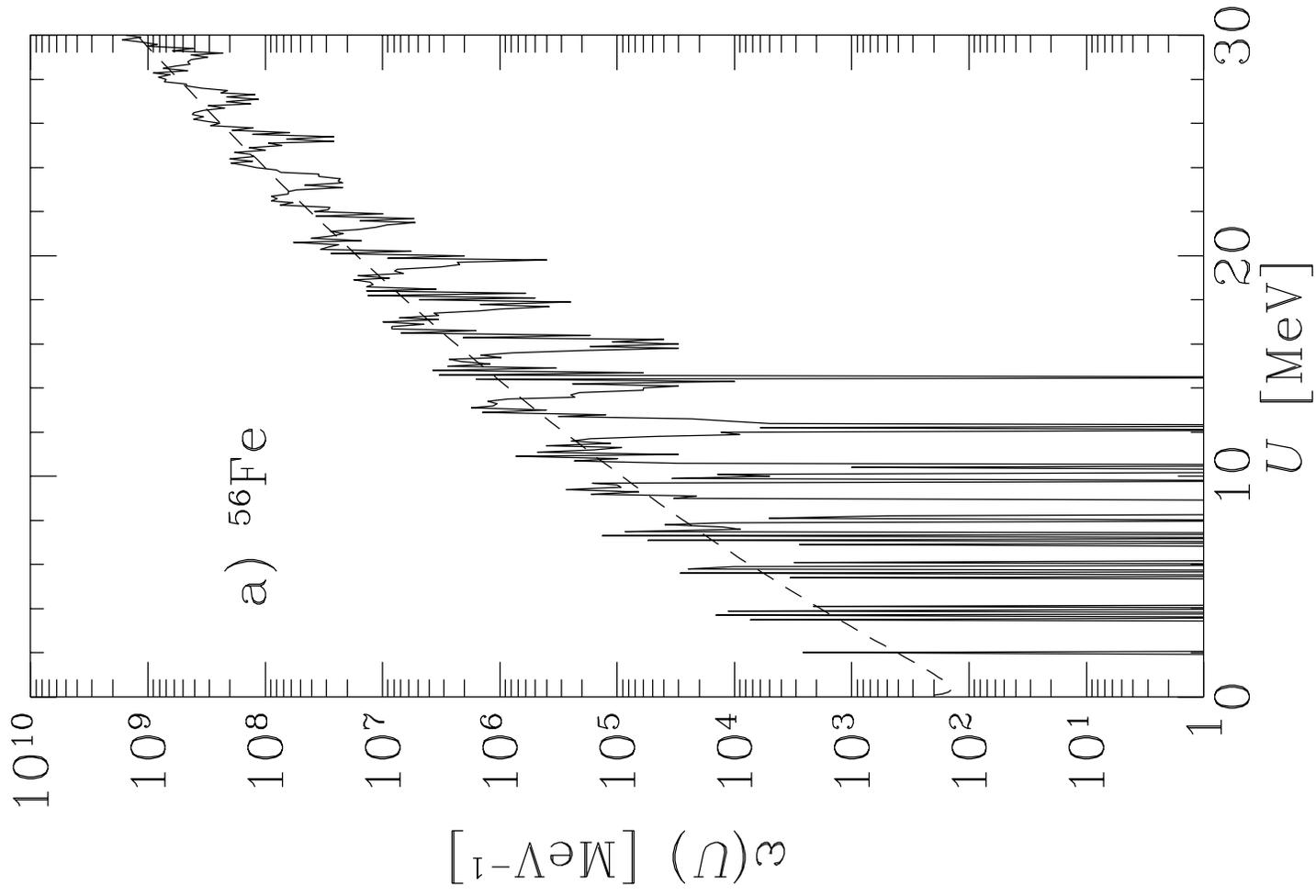
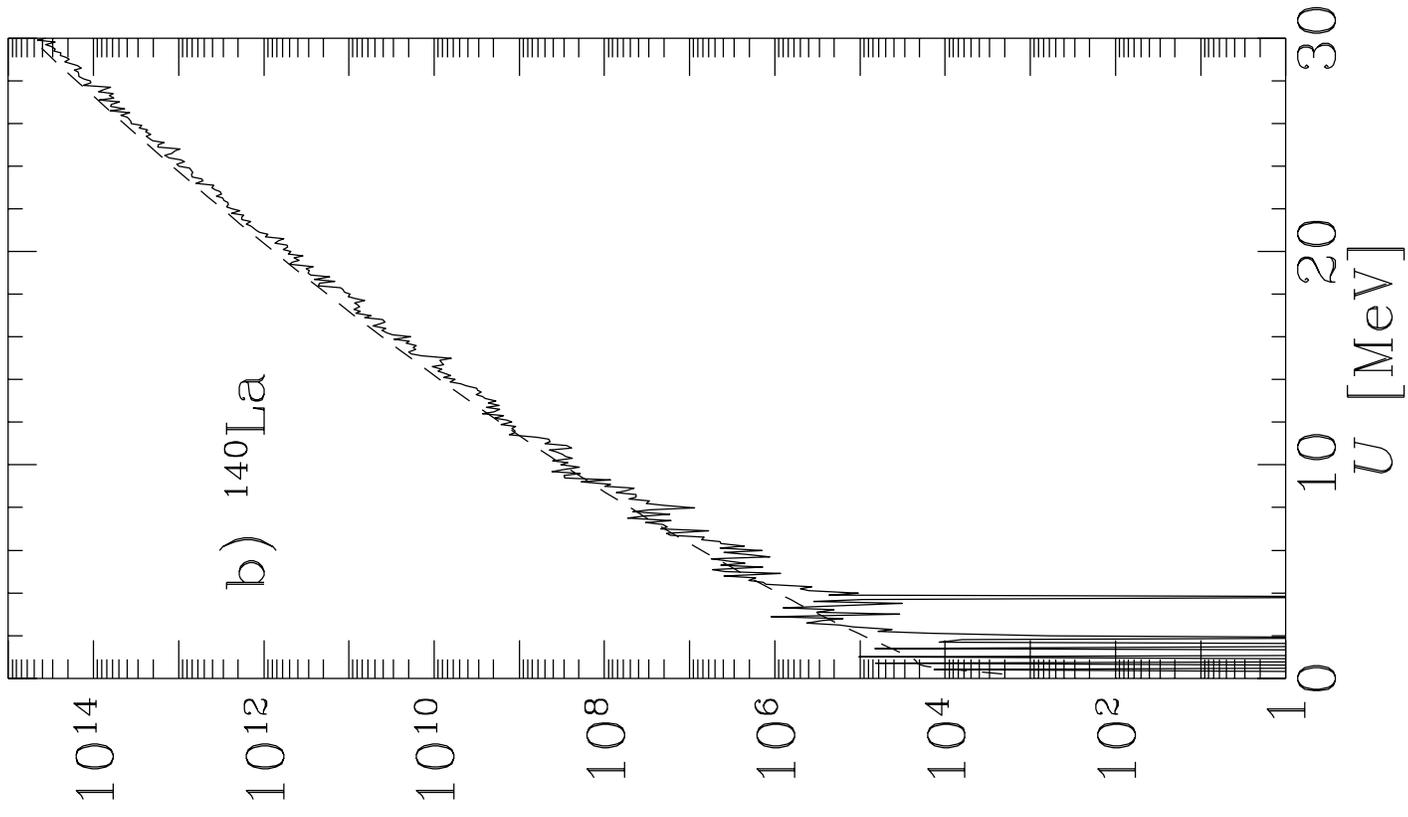
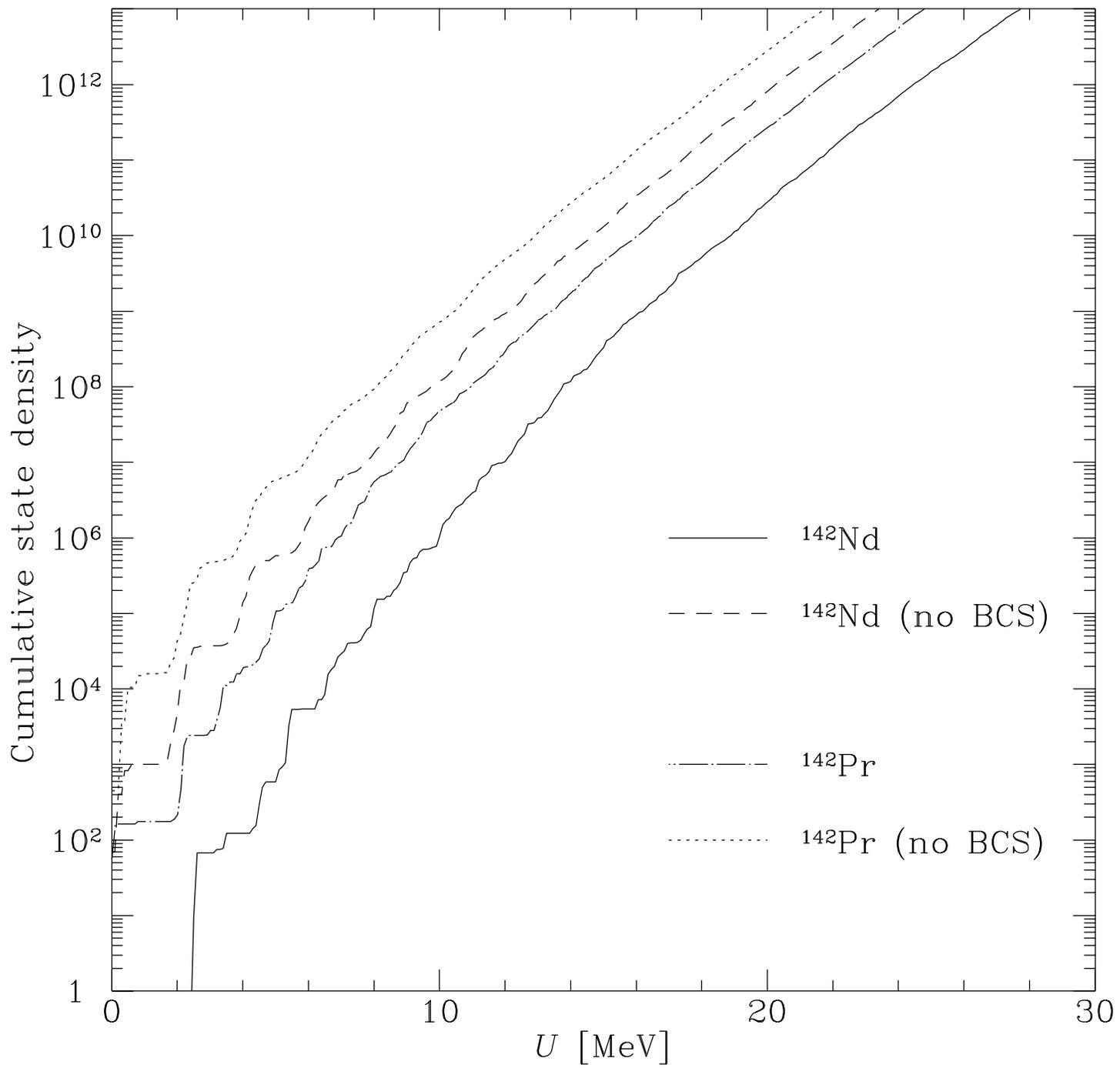
\end{document}